\documentclass[manuscript]{acmart}
\usepackage[utf8]{inputenc}
\AtBeginDocument{%
  }

\setcopyright{acmlicensed}
\copyrightyear{2018}
\acmYear{2018}
\acmDOI{XXXXXXX.XXXXXXX}
\acmConference[Conference acronym 'XX]{Make sure to enter the correct
  conference title from your rights confirmation email}{June 03--05,
  2018}{Woodstock, NY}
\acmISBN{978-1-4503-XXXX-X/2018/06}

\begin{document}

\title{When Readability and Source Retention Diverge: An Evaluability Gap in AI Translation}

\author{Chenchen Mao}
\email{chm321@lehigh.edu}
\affiliation{%
  \institution{Lehigh University}
  \city{Bethlehem}
  \state{Pennsylvania}
  \country{United States}}

\author{Hanjing Shi}
\email{hasa23@lehigh.edu}
\affiliation{%
  \institution{Lehigh University}
  \city{Bethlehem}
  \state{Pennsylvania}
  \country{United States}}

\author{Haiyan Jia}
\email{haj616@lehigh.edu}
\affiliation{%
  \institution{Lehigh University}
  \city{Bethlehem}
  \state{Pennsylvania}
  \country{United States}}

\author{Emily Wegrzyn}
\email{emw224@lehigh.edu}
\affiliation{%
  \institution{Lehigh University}
  \city{Bethlehem}
  \state{Pennsylvania}
  \country{United States}}

\author{Dominic DiFranzo}
\email{djd219@lehigh.edu}
\affiliation{%
  \institution{Lehigh University}
  \city{Bethlehem}
  \state{Pennsylvania}
  \country{United States}}

\renewcommand{\shortauthors}{Mao et al.}

\begin{abstract}
 Readable AI output can leave an evaluability gap: even when the source is shown, an overall-quality judgment may not reflect what an output preserves. We investigated how source-text condition and output rendering relate to perceived translation quality, and how output and system appraisals relate to trust and stated disclosure willingness in a plain-text interface. A focal 2$\times$2 comparison ($N=306$) using TransLingo examined simple generated narratives and complex literary-philosophical prose alongside LLM-generated readability-oriented outputs and researcher-revised fidelity-oriented outputs. A descriptive stimulus audit indicated greater source retention in fidelity-oriented outputs in both source-text conditions. Factorial analyses showed a significant rendering-by-source-text-condition
interaction in perceived quality. Participants rated fidelity-oriented outputs higher than readability-oriented outputs for the simple narratives, whereas no reliable rendering difference emerged for the complex prose. A corresponding source-condition-dependent pattern was observed for perceived intelligence, agency-oriented anthropomorphic attribution, and task-performance trust. A separate theory-ordered appraisal-structure SEM characterized concurrent associations among perceived quality, perceived intelligence, agency-oriented anthropomorphic attribution, task-performance trust, and stated disclosure willingness across six domains, with task-performance trust as the proximal correlate of stated willingness. The observed rating pattern distinguishes source access from source evaluability: for the complex stimuli, displaying the source did not ensure that one overall-quality rating reflected differences in retained content. It also separates support for evaluating translation output from data-handling support for decisions about what personal text to entrust to a system.
\end{abstract}

\ccsdesc[500]{Do Not Use This Code~Generate the Correct Terms for Your Paper}
\ccsdesc[300]{Do Not Use This Code~Generate the Correct Terms for Your Paper}
\ccsdesc{Do Not Use This Code~Generate the Correct Terms for Your Paper}
\ccsdesc[100]{Do Not Use This Code~Generate the Correct Terms for Your Paper}


\maketitle

\section{Introduction}

Machine translation is widely used across many domains, including medicine and legal services \cite{vieira_understanding_2021}, online chat \cite{sahin_multilingual_2014}, and the workplace \cite{bindels_machine_2024}. Translation interfaces are often treated as utilities: a user enters text, receives output, and moves on. Seen this way, trusting a translation system means little more than judging whether the output is usable. But the text that users submit to these systems is often personal — private messages, workplace requests, or material about their own circumstances — and often entered at moments when wording matters and verification is difficult \cite{vieira_understanding_2021}. When the input is about the user themselves, entering it is itself an act of disclosure, and the system becomes not merely an information source but a disclosure interface. Trust then becomes a question of what users are willing to entrust to a system that does not merely transmit their text but may reshape it. This view is reflected in AI-mediated communication research, which treats translation as a case in which a computational agent modifies messages toward a communication goal \cite{HancockNaamanLevy2020AIMC}. For HCI, AI translation is therefore both an output-evaluation problem and an interface through which users entrust information to a computational mediator.

LLM-based translation makes this reformulation especially important. Early machine translation often produced disfluent, awkward output, but contemporary LLMs often generate highly fluent text \cite{ataman_machine_2025}. Fluency alone may therefore no longer reliably distinguish a good rendering from a poor one. What still varies, and what a prompt can shift substantially, is readability: an LLM can be asked to make a text easier to read \cite{pascoal_improving_2026}, but readability gains can come at the cost of faithfulness. Evidence of this tension has emerged across tasks in which LLMs restate source content. In scientific summarization, LLM-generated summaries may read clearly while overgeneralizing the original claims \cite{PetersChinYee2025GeneralizationBias}. In question answering, fluent answers can appear more trustworthy even when they are not faithful to the supporting evidence \cite{chiesurin_dangers_2023}. Similarly, when LLMs are asked to simplify text to a lower reading level, greater degrees of simplification make it more difficult to preserve the original meaning fully \cite{barayan_analysing_2025}. Although these findings come from different tasks, they point to a common concern for LLM-based translation: readability-oriented generation may weaken the model’s adherence to the source. We call the resulting possibility an evaluability gap: users may see the source without being able to recognize, through one overall-quality judgment, what the output retained or altered.

Whether users register the divergence between readability and faithfulness, however, may depend on the source text itself. We therefore treat differences between the tested source-text conditions as an open research question (\textbf{RQ1b}).

Because translation can involve personal text, how users perceive the system may matter alongside how they judge its output. Our setting is a plain-text translation interface without an avatar, voice, persona, or multi-turn conversation. We therefore examine whether users' output evaluations are associated with anthropomorphic or agentic perceptions of the system.

One such system appraisal is perceived intelligence. We therefore examine whether perceived quality is positively associated with perceived intelligence \cite{koda_agents_1996}, and whether perceived intelligence is, in turn, positively associated with perceived anthropomorphism.

The tension between readability and faithfulness, and the possible role of source-text complexity, motivates our overarching question:\begin{quote}
\emph{When translation output is more readable but retains less source content---in an interface with no overt anthropomorphic design cues---how do users evaluate its quality, does the source-text dependence of that evaluation extend to their appraisals of the system itself, and how are those evaluations connected to perceptions of the system, trust, and disclosure willingness?}
\end{quote}

\begin{figure*}[htbp]
    \centering
    \includegraphics[width=\textwidth]{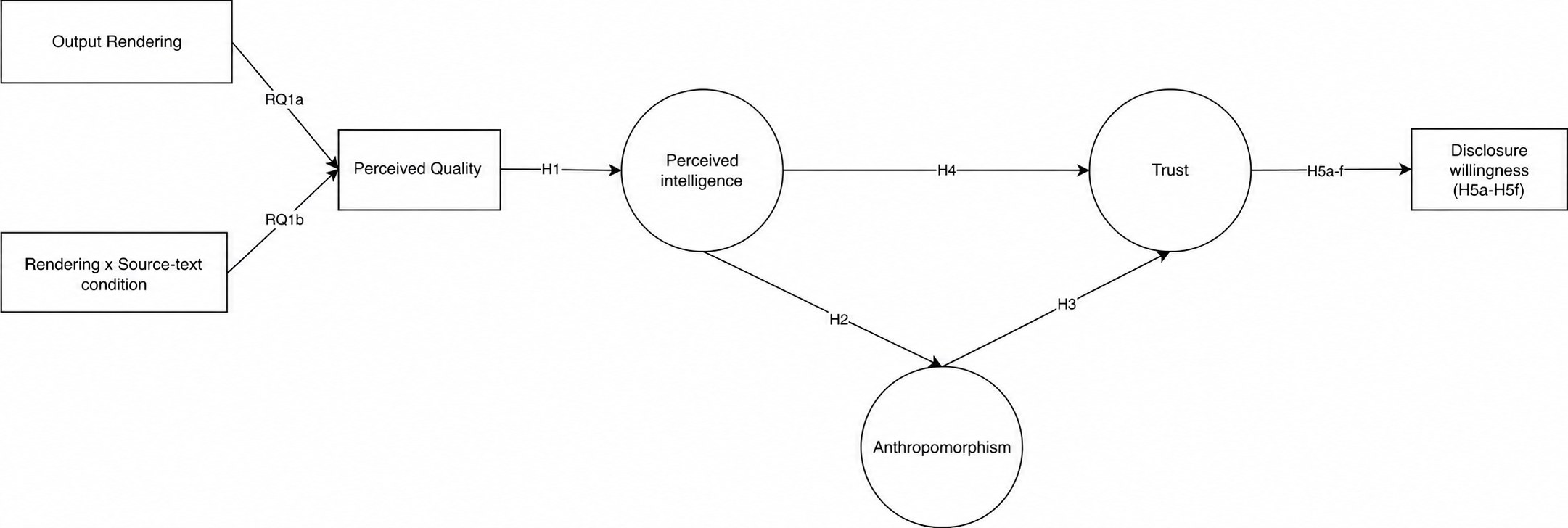}
\caption{Hypothesized model. Output rendering (readability-oriented vs.\ fidelity-oriented)
is compared on perceived quality (RQ1a), with the rendering--quality relationship examined
across the two source-text conditions (RQ1b). Perceived quality is expected to be positively associated
with perceived intelligence (H1); perceived intelligence with perceived anthropomorphism
(H2) and trust (H4); perceived anthropomorphism with trust (H3); and trust with disclosure
willingness across six domains (H5a--H5f): tastes and interests, attitudes and opinions,
work or studies, economic and social status, interpersonal relations and self-concept, and physical
appearance and sex. Paths among the post-task appraisals represent hypothesized associations rather than temporal or causal effects.}
    \Description{Conceptual path diagram. Output rendering and its interaction with source-text condition point to perceived quality. Perceived quality points to perceived intelligence, which points to anthropomorphism and trust. Anthropomorphism also points to trust, and trust points to disclosure willingness in six domains.}
    \label{fig:hyp}
\end{figure*}

We address this question in a focal 2×2 comparison of source-text condition (simple versus complex) and prepared output rendering: readability-oriented output that prioritizes readability and fidelity-oriented output revised toward source fidelity. We first compare perceived quality across the output renderings and ask whether that relationship differs between the source-text conditions (RQ1a, RQ1b). In exploratory analyses, we then ask whether the same rendering-by-source-text pattern appears in users' appraisals of the system itself — perceived intelligence, agency-oriented anthropomorphism, and task-performance trust (RQ2). We then estimate a theoretically ordered pattern of associations among the concurrently measured post-interaction appraisals: perceived quality is associated with perceived intelligence; perceived intelligence is associated with perceived anthropomorphism; perceived intelligence and perceived anthropomorphism are associated with trust; and trust is associated with willingness to disclose across six personal-information domains (H1--H5; Figure~\ref{fig:hyp}).

This study makes three contributions:

\begin{enumerate}
    \item \textbf{It distinguishes source visibility from source evaluability.} Although every participant saw the source and output side by side, the greater source retention documented by the stimulus audit was reflected in overall-quality ratings only for the simple sources. Factorial analyses identified this source-dependent rendering pattern in
perceived quality and showed that it extended to perceived intelligence,
agency-oriented anthropomorphism, and task-performance trust. No reliable rendering contrast emerged for stated disclosure willingness in either source-text condition.

    \item \textbf{It provides a theory-ordered structural account of the appraisal system surrounding translation output.} A complementary appraisal-structure SEM described concurrent associations among perceived quality, perceived intelligence, agency-oriented anthropomorphism, trust, and disclosure willingness. Trust was positively associated with disclosure willingness in every domain. Adding direct anthropomorphism-to-disclosure paths did not improve model fit, supporting the more parsimonious specification in which trust was the appraisal most proximally associated with stated disclosure willingness.

    \item \textbf{It separates output evaluation from information entrustment.} Trust was positively associated with stated disclosure willingness across all six domains (\(\beta=.363\)--\(.566\), all FDR-adjusted \(p<.001\)). In contrast, although the rendering conditions differed on four appraisal measures for the simple sources, they did not differ reliably in stated disclosure willingness. The model-implied associations between rendering and disclosure were correspondingly small for the simple sources (\(\beta=.050\)--\(.078\)) and near zero for the complex sources. This pattern motivates treating support for evaluating an output and support for deciding what personal text to submit as distinct interface-design problems.
\end{enumerate}

These findings expose a limit of source-visible translation interfaces: placing source evidence on screen did not ensure that users' holistic judgments reflected it. Source-linked change information and explicit data-handling information address different decisions and require separate evaluation.

\section{Related Work and Hypothesis Development}
\subsection{When Readability and Source Retention Diverge}

Translation quality is commonly judged through a combination of source adequacy---how faithfully the output preserves the meaning of the source---and target-language fluency \cite{white_evaluation_1993,han_machine_2018}. \citet{martindale_fluency_2018} found that users reacted strongly to disfluent MT but were much less sensitive to fluent output that was not fully adequate. As fluent LLM output becomes common \cite{ataman_machine_2025}, fluency may become less diagnostic among outputs that already read smoothly. Adequacy then remains an important source of quality differences that users may not readily detect.

This is also where adequacy and readability can come apart. An LLM can be prompted to enhance the readability of a text, making it more accessible to particular readers such as students \cite{pascoal_improving_2026}, but rendering a difficult source into simple language can diverge from faithfulness to that source. This tension is not specific to translation; it appears wherever an LLM restates source content for a reader. In scientific summarization, \citet{PetersChinYee2025GeneralizationBias} show that LLM-generated summaries can read as clear and accessible while overgeneralizing the claims of the original papers. Similarly, in conversational question answering, \citet{chiesurin_dangers_2023} find that LLMs can produce fluent, user-aligned answers that raise perceived trustworthiness even when those answers are not faithful to the underlying evidence. These findings raise the possibility of a trade-off between readability and source retention in LLM-generated output.

How users perceive the quality of output that diverges in readability and adequacy in this way, however, is not theoretically settled. Two accounts pull in opposite directions. Research on processing fluency shows that text which is easier to process is judged as more credible and of higher quality, independently of its actual content \cite{ReberSchwarz1999FluencyTruth,AlterOppenheimer2009Fluency} — by this account, a readability-oriented rendering should be perceived as higher in quality. By a competing account, adequacy — how faithfully the output preserves the meaning of the source — has long been recognized as a core dimension of translation quality \cite{white_evaluation_1993,han_machine_2018}; when users can compare the output against the source, a rendering that preserves more of its meaning may be recognized as more adequate, and therefore as higher in quality. By this account, a fidelity-oriented rendering should be perceived as higher in quality. Because these accounts predict opposite effects, we leave the relationship between output rendering and perceived quality as our first open research question:

\textbf{RQ1a}: How do perceived-quality ratings differ between readability-oriented and fidelity-oriented outputs?

The relationship between output rendering and perceived translation quality may differ across source-text conditions. Complex texts require more background knowledge and cognitive effort to understand \cite{wood_task_1986,chan_moderating_2015}. Under such conditions, users may rely more heavily on processing fluency---the subjective ease with which information is understood---as a cue when evaluating quality \cite{AlterOppenheimer2009Fluency}. For example, greater readability has been shown to increase readers' confidence in financial disclosures through experienced processing fluency \cite{rennekamp_processing_2012}. Applied to machine translation, this account suggests that a readability-oriented rendering may be evaluated more favorably when the source text is difficult to process.

Source-text complexity may also make fidelity more difficult to evaluate. When users struggle to understand the source itself, they may be less able to determine whether a translation accurately preserves its meaning. The relative advantage of a more faithful rendering may therefore become less apparent as source-text complexity increases. Both mechanisms imply that complexity may increase the importance of readability while making fidelity harder to recognize. However, it remains unclear whether either mechanism is strong enough to alter users' overall quality judgments, and our study does not measure the mechanisms separately. We therefore examine whether the rendering--quality relationship differs across the tested source-text conditions:

\textbf{RQ1b}: Does the relationship between output rendering and perceived translation quality differ across the tested source-text conditions?

\subsection{From Perceived Quality to Perceived Intelligence}

Beyond how users judge the translation itself, output quality may also relate to how they perceive the system behind it. In this study, we adopt the definition by \citet{Moussawi_Koufaris_2019}, characterizing perceived intelligence as the extent to which an agent appears efficient, useful, goal-directed, and autonomous, with the specific ability to generate effectual outputs and process natural language. Perceived translation quality concerns users' evaluation of the specific outputs presented to them, whereas perceived intelligence concerns the broader system capabilities they infer from those outputs.

Prior research suggests that users draw on a system's observable performance when judging its intelligence. \citet{koda_agents_1996} found that the perceived intelligence of a poker-playing agent depended more on its demonstrated competence than on its facial appearance. Similarly, \citet{10.3389/fpsyg.2023.1129294} experimentally varied the reliability of in-vehicle agents and found that highly reliable agents were perceived as more intelligent than less reliable agents. Consistent with these findings, a review of human--robot interaction research identified successful task performance as one factor commonly associated with higher perceived intelligence \cite{tusseyeva_perceived_2024}.

In our setting, the most directly observable aspect of system performance is the perceived quality of its translation output. We therefore expect perceived translation quality to be positively associated with how intelligent the system is perceived to be.

\textbf{H1}: Higher perceived translation quality is positively associated
with higher perceived intelligence of the system.

\subsection{Anthropomorphism: Beyond Designed Social Cues}

Judging a system as intelligent is not only an assessment of what it can do; it may also be associated with how the system is categorized—from a tool that processes text to an agent that appears to understand and reshape it. Prior work describes this humanlike interpretation as anthropomorphism, broadly characterized as the attribution of human traits and characteristics to computers and other interactive technologies \cite{nass_machines_2000}. In the present study, we focus on an agency-oriented form of anthropomorphism: perceiving a nonhuman system as possessing humanlike capacities to understand, anticipate, and plan \cite{gray_dimensions_2007,waytz_mind_2014}.

This expectation is consistent with the three-factor theory of anthropomorphism \cite{epley_seeing_2007}, which identifies elicited agent knowledge, effectance motivation, and sociality motivation as three determinants of anthropomorphism. Elicited agent knowledge is particularly relevant here: it refers to the activation of knowledge ordinarily used to understand human agents. When a system appears capable of understanding, anticipating, and planning, users may draw on this knowledge and interpret the system in more humanlike terms. Perceived intelligence may therefore be positively associated with perceived anthropomorphism even when the interface contains no explicitly designed humanlike cues.

Much prior research has examined anthropomorphism in interfaces containing designed social cues, including humanlike appearance, names, voices, and behavioral features \cite{waytz_mind_2014,nowak_influence_2005,salem_err_2013}. This includes studies reporting positive associations between perceived intelligence and anthropomorphism \cite{troshani_we_2021,Moussawi_Koufaris_2019,moussawi_how_2021,moussawi_effect_2021,ma_effect_2025}. Because social cues were also present in these settings, it remains unclear whether the association holds when the system's textual output is the primary basis for evaluation.

\textbf{H2}: Higher perceived intelligence of the system is positively associated with greater perceived anthropomorphism.

\subsection{Anthropomorphism, Perceived Intelligence, and Trust}

Perceiving a system as more agent-like may also be associated with how far users are willing to rely on it. Experimental work has manipulated anthropomorphic cues and found effects on trust. \citet{waytz_mind_2014} varied an autonomous vehicle's humanlike features and found greater trust when those features were present, while \citet{natarajan_effects_2020} identified anthropomorphism as one of the strongest predictors of trust and compliance in a multi-factor robot study. Survey evidence reports a similar positive association \cite{gomes_anthropomorphism_2025}. Focus-group research further suggests that consumers regard humanness and intelligence as important bases for judging whether an AI service can be trusted \cite{troshani_we_2021}.

\textbf{H3}: Greater perceived anthropomorphism is positively associated with higher trust in the system.

Perceived intelligence may also be associated with trust alongside its association with perceived anthropomorphism. In a survey of employee beliefs about AI, \citet{lu_role_2025} found that trust mediated the relationship between perceived intelligence and acceptance of AI use. \citet{chen_thinking_2025} found that perceived intelligence was positively associated with both cognitive and emotional trust. Focus-group evidence similarly indicates that consumers rely on perceived intelligence and anthropomorphism as key determinants of AI trustworthiness \cite{troshani_we_2021}.

\textbf{H4}: Higher perceived intelligence is positively associated with higher trust in the system.

\subsection{Trust and Disclosure Willingness}

The related work and hypotheses above concern the appraisals associated with trust in a translation system. The practical significance of trust lies in users' willingness to provide personal text to the system \cite{jourard1971self}.

Research on privacy decision-making offers an account of what makes users willing to do so. The extended privacy-calculus model of \citet{dinev_extended_2006} conceptualizes disclosure as a decision under uncertainty, in which privacy concerns inhibit online transactions while Internet trust and personal interest can together outweigh perceived privacy risks. In this model, trust is one factor that can make disclosure more likely. Supporting this role, \citet{metzger_privacy_2004} identified trust as an important antecedent of disclosure to commercial websites, while \citet{krasnova_online_2010} found that trust in a network provider reduced the perceived risk associated with self-disclosure.

These findings come from e-commerce sites and social platforms, but the same decision arises whenever users provide personal text to a system---including a translation interface. We therefore expect trust in the translation system to be positively associated with disclosure willingness, and because that willingness may vary with the topic being disclosed, we examine the association separately across six personal-information domains.

\textbf{H5a--H5f}: Higher trust in the system is associated with greater willingness to self-disclose across each of the six personal-information domains: tastes and interests, attitudes and opinions, work or studies, economic and social status, interpersonal relations and self-concept, and physical appearance and sex.

Because these appraisals were collected after participants viewed the same prepared outputs, we also ask, as an exploratory question, whether the source-condition dependence examined for perceived quality (RQ1a,RQ1b) extends to the appraisals of the system itself:

\textbf{RQ2}: Do the relationships between output rendering and perceived intelligence, perceived anthropomorphism, and task-performance trust also differ across the tested source-text conditions?

\section{Methods}

\subsection{Experiment Design}
The experiment contained a focal $2 \times 2$ comparison. The first factor was source-text condition, contrasting simple and complex texts. The second factor was output rendering, comparing a readability-oriented rendering---produced by an LLM and oriented toward making the output easy to read---with a fidelity-oriented rendering of that same output, revised toward closer source fidelity. All outputs were presented as TransLingo translations; the prepared-output contrast changed the text rather than adding a visible social cue.

\subsection{Selection and Preparation of Source Texts and Translations}

\subsubsection{Source Texts} \label{3.2.1}
To construct the source-text conditions, we varied the source text because its characteristics shape the complexity of the translation task. We based our approach on the work of \citet{wood_task_1986}, who argues that task complexity includes both component and coordinative complexity. This concept was recently extended to Human-AI decision-making contexts by \citet{salimzadeh_missing_2023}. We operationalized component complexity---the number of distinct information cues to be processed---through vocabulary; simple texts contained common, frequently repeated words (e.g., village, flower, children), while complex texts featured abstract, low-frequency words (e.g., paternal affection, solecism, steadfastness). We addressed coordinative complexity---the nature of relationships between task inputs and outputs---through sentence structure, using short sentences averaging 15 words for simple texts and long sentences averaging 72 words for complex texts. The simple texts were generated for this study using ChatGPT-4 with instructions to create content easily understood by primary school students. In contrast, the complex texts were extracted directly from Aurelius's \textit{Meditations} \cite{aurelius1923thoughts}. We assessed participants' perceptions of this contrast using the study-specific grade-level item described below.

Two source texts were used at each complexity level, four in total. Each participant was presented with both texts at their assigned complexity level and rated each one separately, so every participant contributed two readability judgments and two translation-quality judgments. Because each complexity level is represented by texts drawn from a single provenance, source complexity is confounded with genre, topic, provenance, period style, and length: the simple texts are contemporary LLM-generated narrative prose, whereas the complex texts are drawn from a nineteenth-century English translation of Stoic philosophy \cite{aurelius1923thoughts}. The source-text comparison should therefore be read as a contrast between simple generated narrative and complex literary-philosophical prose rather than as a pure manipulation of linguistic complexity.

To assess the intended source-text contrast, participants rated the level of education required to read and understand the input texts using a measure adapted from the Dale--Chall Readability Formula \cite{chall_readability_1995}. The simple texts were rated as accessible to an average seventh- or eighth-grade student, whereas the complex texts were rated as requiring approximately college-level comprehension. A Mann--Whitney $U$ test showed a significant difference between the two conditions, $U=607.50$, $Z=-18.20$, $p<.001$, supporting the intended contrast in perceived source-text complexity.

\subsubsection{Output Rendering Condition} \label{3.2.2}

Output rendering had two levels: a readability-oriented rendering and a fidelity-oriented rendering. For the readability-oriented rendering condition, ChatGPT-4 paraphrased the source text in a way that prioritized making the final English output easy to read, while preserving the general topic of the source. For the fidelity-oriented rendering condition, a researcher revised the readability-oriented output for closer faithfulness to the source — restoring content that had been simplified or omitted, while keeping the output readable.

To assess whether the two rendering conditions differed in the intended directions on readability and source retention, we conducted a stimulus-level audit on the source--output pairs used in the study. Each pair was scored on four metrics: the change in Flesch reading-ease score from source to output ($\Delta$Flesch), which captures readability gain \cite{kincaid1975derivation}; content-word Jaccard overlap, which captures how much source content survives in the output; BERTScore F1, which captures semantic similarity to the source \cite{Zhang2020BERTScore}; and COMET, a learned metric for translation quality \cite{rei2020comet}.

\begin{table}[h!]
\centering
\resizebox{\columnwidth}{!}{%
\begin{tabular}{llcccc}
\toprule
\textbf{Source} & \textbf{Rendering} & \textbf{$\Delta$Flesch} & \textbf{Jaccard} & \textbf{BERTScore} & \textbf{COMET} \\
\midrule
Simple & Readability-oriented & 14.08 & 0.377 & 0.937 & 0.869 \\
Simple & Fidelity-oriented & 4.83 & 0.713 & 0.976 & 0.914 \\
Complex & Readability-oriented & 70.43 & 0.138 & 0.854 & 0.670 \\
Complex & Fidelity-oriented & 56.40 & 0.452 & 0.910 & 0.753 \\
\bottomrule
\end{tabular}}
\caption{Stimulus-level audit of the prepared-output contrast across the four focal cells. $\Delta$Flesch = change in Flesch reading-ease score from source to output; Jaccard = content-word Jaccard overlap with source; BERTScore = BERTScore F1 against source; COMET = learned MT quality metric. Higher $\Delta$Flesch indicates a larger readability gain; higher Jaccard and BERTScore indicate closer source retention, while higher COMET indicates higher estimated MT quality.}
\label{tab:stim-audit}
\end{table}

Table~\ref{tab:stim-audit} summarizes differences among the prepared outputs. First, the readability-oriented outputs made the source texts easier to read. Their reading-ease gains were larger than those of the fidelity-oriented outputs for both simple texts ($14.08$ vs.\ $4.83$) and complex texts ($70.43$ vs.\ $56.40$). Second, compared with the readability-oriented outputs, the fidelity-oriented outputs retained a larger proportion of the source content words (Jaccard), had higher semantic similarity to the source (BERTScore), and received higher estimated translation-quality scores (COMET) at both complexity levels. Thus, both conditions produced readable English, but they emphasized different goals: the readability-oriented condition simplified the source more aggressively, whereas the fidelity-oriented condition retained more of its original content. These metrics describe relative differences among the stimuli rather than independently validated translation adequacy.

\subsection{Experiment Procedure}
Participants first completed a pre-survey reporting their demographics, first language, and other languages spoken; language proficiency was not assessed using a standardized test. They were then introduced to TransLingo, a prototype AI-mediated translation system. Data were collected in two separate periods: one evaluated the readability-oriented rendering and the other evaluated the fidelity-oriented rendering. Within each period, participants were randomly assigned to either the simple or complex source-text condition, resulting in a 2 × 2 design defined by rendering and source-text complexity. Each participant evaluated two texts in the assigned complexity condition. Thus, source-text complexity was randomized within each period, whereas the rendering comparison was conducted between periods, as discussed further in the Limitations.

Inside TransLingo, participants were told that two test texts had been randomly picked from the internet. Because the sample was English-speaking, the task translated English to Chinese and then back to English, allowing participants to evaluate the final English output.

After reviewing both the source and translated texts in their assigned condition, the post-survey button became active. The post-survey measured perceived input-text complexity, perceived translation quality, perceived intelligence, anthropomorphism, trust in the system, and willingness to disclose personal information while using the system.

\subsubsection{Ethical Consideration}
The design involved mild deception: participants were told they were testing a working prototype, whereas all input texts and translations had been prepared in advance and held constant within condition. This was necessary to hold stimulus quality fixed rather than allowing it to vary with live model output, and to maintain experimental realism. At the end of the study, participants were debriefed and informed that the materials had been selected or prepared in advance, why this information had been withheld beforehand, and whom to contact with questions or to withdraw. The study procedures received approval from the Institutional Review Board (IRB) at Anonymous University.

\subsection{TransLingo Interface Overview}
TransLingo was the web interface used in the experiment. Participants selected the pre-made source text assigned by the interface, clicked the 'translate' button, and then saw the source text and translated output together.

\begin{figure}
\centering
\includegraphics[width=1\linewidth]{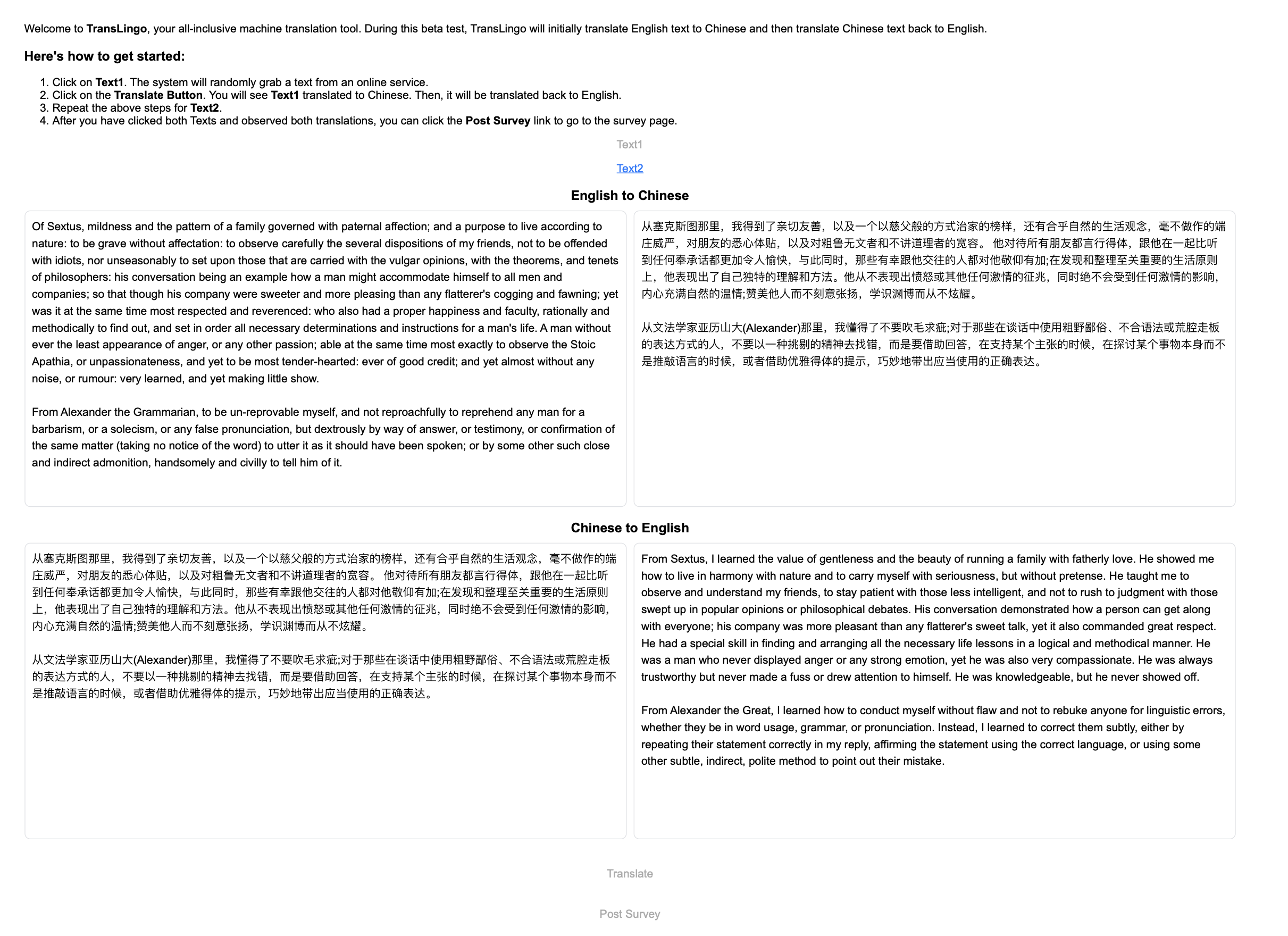}
\caption{Snapshot of the TransLingo interface used in the experiment. Participants saw the English source, an intermediate Chinese text, and the final English output.}
\Description{Screenshot of the TransLingo prototype. The interface presents an English source-text panel, a Chinese intermediate-translation panel, and an English back-translation panel.}
\label{fig:interface}
\end{figure}

As illustrated in Figure~\ref{fig:interface}, the interface presented the English source, intermediate Chinese text, and final English output in separate panels.

\subsection{Variables and Measurements} \label{3.5}

All post-survey constructs were measured after participants had reviewed both source texts and both translated outputs in their assigned condition.

For the factorial analyses, item responses for perceived intelligence, anthropomorphism, and trust were averaged into composite scores; in the appraisal-structure SEM, the same items were modeled as indicators of their respective latent factors.

\subsubsection{Perceived source-text readability (manipulation check)}

For each of the two source texts, participants answered "What level of education is needed for someone to read and understand this text?" on a 7-point scale adapted from grade-level categories associated with the Dale–Chall Readability Formula \cite{chall_readability_1995}, ranging from 1 ("easily understood by an average 4th-grade student or lower") to 7 ("easily understood by an average college graduate student and above"). The two ratings were averaged. This study-specific subjective item was used as a manipulation check, not as an objective application of the formula or a validated psychometric scale.

\subsubsection{Perceived translation quality}

For each of the two texts, participants were shown the source text together with the final English output in their assigned rendering condition and answered: "Please rate the overall quality of the translation compared to the original source text provided above." Responses used a 7-point scale labeled Extremely poor (1), Poor (2), Fair (3), Average (4), Good (5), Very good (6), and Excellent (7). Perceived quality was measured with a single item per text; the two ratings were strongly correlated, $r = .826$, and were averaged into an observed composite ($M = 5.37$, $SD = 1.18$; Spearman–Brown coefficient $= .904$). Perceived quality entered the structural model as an observed composite rather than as a latent variable because it was a single item administered twice, not a multi-item scale. The Spearman–Brown coefficient reflects consistency across the two text-level ratings; the item does not identify whether participants weighted readability, adequacy, or another quality dimension.

\subsubsection{Perceived intelligence}

Perceived intelligence was measured using four items adapted from \citet{Moussawi_Koufaris_2019}. References to the original conversational agent and its task were replaced with TransLingo and machine-translation-specific language. For example, participants rated the statement, ``The machine translation system (TransLingo) can understand the input text accurately.'' Responses were recorded on a 7-point scale from 1 (``strongly disagree'') to 7 (``strongly agree''). The four items were modeled as indicators of a latent perceived-intelligence factor ($\alpha=.914$, $CR=.915$, $AVE=.730$).

\subsubsection{Anthropomorphism}

Anthropomorphism was measured using four items adapted from \citet{waytz_mind_2014}. The system referent and task context were rewritten for machine translation while retaining the capacities assessed by the source measure. For example, participants were asked, ``How well do you think this machine translation system (TransLingo) can plan and choose the best translation
options available?'' Responses used the original 0--10 format, with higher values indicating greater attribution of the specified capacity. 

The four items were modeled as indicators of a latent anthropomorphism factor ($\alpha=.844$, $CR=.858$, $AVE=.606$). This operationalization primarily captures cognitive agency---including perceived understanding, anticipation, and planning---rather than humanlike appearance, personality, or emotional experience. 

\subsubsection{Trust}

Trust was measured using seven items adapted from the self-reported trust measure reported by \citet{waytz_mind_2014}. Vehicle-specific referents and performance scenarios were replaced with TransLingo and translation-related tasks. For example, participants were asked, ``How much do you trust this machine translation system's (TransLingo) ability to provide accurate translations in the next task?'' Responses used the original 0--10 format, with higher scores indicating greater trust. 

The seven items were modeled as indicators of a latent trust factor ($\alpha=.953$, $CR=.956$, $AVE=.755$). Because the items concern translation accuracy, reliability, and task performance, the construct represents performance-related trust in TransLingo rather than trust in its privacy or data-handling practices.

\subsubsection{Disclosure willingness}

Disclosure willingness was measured using a 36-item battery adapted from \citet{10.1145/2858036.2858414}, building on earlier research on self-disclosure \cite{Jourard_Lasakow_1958,rubin_friendship_1978}. The original social-media context was replaced with the machine-translation context. For every topic, participants were asked:

\begin{quote}
``How comfortable do you feel using TransLingo to translate and disclose this aspect of yourself?''
\end{quote}

For example, one item concerned participants' favorite foods and preferred ways of preparing food. Responses ranged from 1 (``very uncomfortable'') to 4 (``very comfortable''). 

The battery contained six topics within each of six domains: tastes and interests, attitudes and opinions, work or studies, economic and social status, interpersonal relationships and self-concept, and physical appearance and sex. Following the instrument's content-sampling structure, the six responses within each domain were averaged to form a domain score. The resulting six domain means were used as separate observed outcome scores in the factorial analyses and were entered into the appraisal-structure SEM as six separate observed endogenous outcomes rather than as indicators of a single disclosure factor. Higher scores indicate greater comfort with disclosing information from the corresponding domain while using TransLingo. Internal consistency was high across the six domain composites ($\alpha=.887$--$.926$).
\subsubsection{Study Variables and Numerical Coding}

The analyses covered the four cells of the focal comparison:
simple/readability-oriented, simple/fidelity-oriented,
complex/readability-oriented, and complex/fidelity-oriented.

For the factorial ANOVAs, output rendering and source-text condition were
entered as two-level factors using sum-to-zero contrasts. This coding allowed
the Type III tests to estimate marginal main effects averaged over the other
factor.

For the appraisal-structure SEM, output rendering was dummy-coded as 0 =
readability-oriented and 1 = fidelity-oriented, and source-text condition was
coded as 0 = simple and 1 = complex. Their interaction was calculated as the
product of the two dummy-coded variables. Under this coding, the rendering
coefficient represents the rendering contrast within the simple-source
reference condition, whereas the source-condition coefficient represents the
source contrast within the readability-oriented reference condition.
\subsection{Data Analysis Methods}
\label{dataana}

The primary analyses used the retained 2$\times$2 focal sample. We conducted
two complementary analyses that addressed different questions.

First, condition differences were examined using separate 2$\times$2
factorial ANOVAs for perceived quality, perceived intelligence,
anthropomorphism, task-performance trust, and each of the six disclosure
domains. Perceived quality was represented by the mean of the two text-level
ratings. Perceived intelligence, anthropomorphism, and trust were represented
by their respective composite means for these factorial analyses, and each
disclosure outcome was represented by the mean of its six constituent items.
Each ANOVA included output rendering, source-text condition, and their
interaction.

Because the four cells were unequal in size, omnibus effects were evaluated
using Type III tests with sum-to-zero contrasts. Planned rendering contrasts
within each source-text condition were estimated using HC3
heteroskedasticity-robust standard errors and confidence intervals. Hedges'
$g$ was used as the standardized effect size for these simple contrasts.
Perceived quality was the single focal outcome corresponding to RQ1 and was
not multiplicity-adjusted. Benjamini--Hochberg false-discovery-rate correction
was applied across the three system-appraisal interaction tests and separately
within each six-test family for the disclosure outcomes. HC3-robust omnibus
tests were used as robustness checks for the Type III interaction tests.

Because each source-text condition was represented by two texts, the rendering
contrasts were additionally estimated separately for each text using
HC3-robust inference. For perceived quality, a long-form three-way diagnostic included both
text-level ratings from each participant and used participant-clustered
robust standard errors.

Second, we estimated an appraisal-structure SEM to characterize the
theory-ordered concurrent associations among perceived quality, perceived
intelligence, anthropomorphism, task-performance trust, and stated disclosure
willingness. Perceived intelligence, anthropomorphism, and trust were modeled
as latent factors using their item-level indicators. Perceived quality was
entered as an observed composite of the two text-level ratings. The six
disclosure domains were represented by separate observed domain means and
were modeled simultaneously as endogenous outcomes, with their residual
covariances freely estimated.

The SEM specified associations from output rendering, source-text condition,
and their interaction to perceived quality; from perceived quality to
perceived intelligence (H1); from perceived intelligence to anthropomorphism
(H2) and trust (H4); from anthropomorphism to trust (H3); and from trust to
stated disclosure willingness across the six domains (H5a--H5f). Because the
post-task appraisal constructs were measured concurrently, the paths among
them were interpreted as theory-ordered associations rather than causal or
temporal effects.

The SEM was estimated in R using the \texttt{lavaan} package
\cite{zis-Rosseel2012Lavaan}, with MLR estimation and full-information maximum
likelihood for missing data. No values were missing among the variables used
in the retained analytic sample. Indirect associations were tested using
5{,}000-sample percentile-bootstrap confidence intervals with fixed random
seeds. Because \texttt{lavaan} does not combine nonparametric bootstrapping
with MLR standard errors, bootstrap intervals were obtained from ordinary
maximum-likelihood refits of the same model. Benjamini--Hochberg correction
was applied within each six-effect family across the disclosure domains.
\subsection{Recruitment and Participants}

An a priori power analysis was conducted in G*Power for a fixed-effects 2 × 2 factorial ANOVA, focusing on the one-degree-of-freedom interaction. Assuming a medium effect size (Cohen’s ($f=.25$)), $(\alpha=.05)$, and 95\% statistical power, the analysis indicated a minimum required sample size of 210 participants. The final analytic sample ($N=306$) exceeded this requirement; the calculation applies to the rendering-by-source interaction rather than to the SEM or indirect associations. Data were collected through Prolific between August 2023 and February 2025. Participants received \$6.00 for completing the study. Responses were screened for data quality, attention-check performance, valid experimental-condition coding, and reported language background. Two respondents whose reported language information indicated Chinese were excluded. The final analytic sample comprised 306 participants across the focal $2 \times 2$ comparison: simple/readability-oriented ($n = 80$), complex/readability-oriented ($n = 88$), simple/fidelity-oriented ($n = 68$), and complex/fidelity-oriented ($n = 70$).

Participants self-identified as 155 women (50.7\%), 141 men (46.1\%), 8 non-binary or third gender participants (2.6\%), and 2 who preferred not to say (0.7\%). After one invalid age response was treated as missing, ages ranged from 18 to 72 ($M=36.6$, $SD=12.5$). The largest education groups were bachelor's degree (40.8\%), some college (19.6\%), high school graduate (14.1\%), master's degree (9.2\%), and associate degree (8.8\%). Race and ethnicity were collected as a multi-select item; the most frequent selections were White or Caucasian (73.2\%), Black or African American (16.7\%), Hispanic, Latino/a, or Spanish Origin (10.5\%), and Asian (8.2\%). Most participants reported English as their first language (304, 99.3\%), and 44 (14.4\%) reported fluency in at least one additional language. Demographic balance checks found no detectable differences across cells in age, gender, race/ethnicity, education, or bilingual status (all $p>.05$).

\section{Findings}

We first report the measurement properties of the latent constructs, followed
by the factorial condition analyses and the appraisal-structure SEM.

\subsection{Measurement Model and Discriminant Validity}

The three-factor confirmatory factor analysis of perceived intelligence, anthropomorphism, and trust showed adequate fit, robust $\chi^2(87) = 223.59$, $p < .001$, CFI $= .939$, TLI $= .927$, RMSEA $= .072$, 90\% CI $[.062, .081]$, SRMR $= .038$.  All standardized loadings were significant and ranged from $.577$ to $.937$. Reliability and convergent validity met conventional benchmarks for all three constructs: perceived intelligence ($\alpha = .914$, $CR = .915$, $AVE = .730$), anthropomorphism ($\alpha = .844$, $CR = .858$, $AVE = .606$), and trust ($\alpha = .953$, $CR = .956$, $AVE = .755$).

Because the appraisal constructs were substantially intercorrelated, we assessed their discriminant validity using HTMT. The HTMT values were $.617$ between perceived intelligence and anthropomorphism, $.705$ between perceived intelligence and trust, and $.871$ between anthropomorphism and trust. The last value is below the $.90$ criterion but above the more conservative $.85$ threshold. As an additional test, constraining the latent correlation between anthropomorphism and trust to unity significantly worsened model fit, robust $\Delta\chi^2(1)=19.48$, $p<.001$. These results support treating anthropomorphism and trust as distinct but closely related constructs.

\subsection{Factorial Condition Differences Across Outcomes}
\label{sec:experimental-total-effects}

We used 2$\times$2 factorial ANOVAs to test whether the rendering contrast
differed between the simple- and complex-source conditions. Table~\ref{tab:condition-results}
reports the cell descriptives and interaction tests.

\subsubsection{Perceived Quality (RQ1a and RQ1b)}
\label{sec:quality-effects}

The rendering contrast depended on source-text condition,
$F(1,302)=4.51$, $p=.034$, $\eta_p^2=.015$ (RQ1b). For the simple sources,
the fidelity-oriented condition received higher quality ratings than the
readability-oriented condition, $\Delta M=.584$, 95\% CI $[.248,.921]$,
$p<.001$, Hedges' $g=.567$ (RQ1a). For the complex sources, the two
rendering conditions did not differ, $\Delta M=.018$, 95\% CI
$[-.376,.413]$, $p=.927$, $g=.014$.

The simple-source contrast was significant for both Text~1,
$\Delta M=.563$, 95\% CI $[.217,.909]$, $p=.001$, and Text~2,
$\Delta M=.605$, 95\% CI $[.231,.979]$, $p=.002$, whereas neither
complex-source contrast was significant (both $p\geq.818$). A direct
three-way test found no evidence that the rendering-by-source-condition
interaction differed between the two texts, $b=-.099$, 95\% CI
$[-.427,.230]$, $p=.556$. The pattern was therefore consistent across the
stimuli employed, although using only two texts at each complexity level
limits generalization beyond them.

\subsubsection{System Appraisals (RQ2)}

The same rendering-by-source-condition pattern appeared for perceived
intelligence, anthropomorphism, and task-performance trust. The interactions
were significant for perceived intelligence, $F(1,302)=5.66$, $p=.018$,
$\eta_p^2=.018$, FDR-adjusted $q=.027$; anthropomorphism,
$F(1,302)=11.56$, $p<.001$, $\eta_p^2=.037$, $q=.002$; and trust,
$F(1,302)=4.20$, $p=.041$, $\eta_p^2=.014$, $q=.041$. The trust
interaction was small and close to the conventional significance threshold
and should therefore be interpreted cautiously.

For the simple sources, the fidelity-oriented condition received higher
ratings of perceived intelligence, $\Delta M=.404$, 95\% CI
$[.146,.662]$, $p=.002$, $g=.489$; anthropomorphism, $\Delta M=1.051$,
95\% CI $[.401,1.701]$, $p=.002$, $g=.521$; and trust, $\Delta M=.730$,
95\% CI $[.132,1.329]$, $p=.017$, $g=.396$. No reliable rendering differences emerged for these outcomes under the complex sources (all \(p\geq .119\)). The point estimate for anthropomorphism was negative in this condition (\(\Delta M=-.497\), 95\% CI \([-1.124,.129]\), \(p=.119\), Hedges’ \(g=-.253\)), but the confidence interval included zero.  Unlike perceived quality, perceived intelligence, anthropomorphism, and trust
were each rated once at the system level after participants had reviewed both
texts. Consequently, text-specific rendering contrasts could not be estimated
for these outcomes.

Averaged across rendering conditions, perceived intelligence was lower in
the complex-source condition than in the simple-source condition,
$F(1,302)=15.98$, $p<.001$, $\eta_p^2=.050$. This marginal difference
should be interpreted alongside the significant interaction, which indicates
that its magnitude varied across rendering conditions.

\subsubsection{Disclosure Willingness}

No rendering-by-source-condition interaction was significant for any of the
six disclosure domains (all $p\geq.076$; all FDR-adjusted $q\geq.360$).
Within the simple-source condition, the unadjusted contrasts reached
$p<.05$ for work or studies, $\Delta M=.243$, $p=.042$, and economic and
social status, $\Delta M=.292$, $p=.037$, but neither remained significant
after correction across the six domains (both $q=.126$). No rendering
contrast was significant within the complex-source condition
(all $p\geq.526$). Thus, no reliable condition difference was detected for
stated disclosure willingness. Across all outcomes, HC3-robust omnibus tests reproduced the same pattern of
significant and nonsignificant rendering-by-source-condition interactions as
the Type III tests.

\begin{table*}[htbp]
\centering
\caption{Condition descriptives and rendering-by-source-condition interaction
tests.}
\label{tab:condition-results}
\resizebox{\textwidth}{!}{%
\begin{tabular}{lccccrrrr}
\toprule
\textbf{Outcome}
& \textbf{\shortstack{Simple\\readability\\$M$ ($SD$)}}
& \textbf{\shortstack{Simple\\fidelity\\$M$ ($SD$)}}
& \textbf{\shortstack{Complex\\readability\\$M$ ($SD$)}}
& \textbf{\shortstack{Complex\\fidelity\\$M$ ($SD$)}}
& \textbf{$F$}
& \textbf{$p$}
& \textbf{$\eta_p^2$}
& \textbf{FDR $q$} \\
\midrule

Perceived quality
& 5.269 (1.003)
& 5.853 (1.051)
& 5.210 (1.364)
& 5.229 (1.141)
& 4.51 & $.034^{*}$ & .015 & --- \\

Perceived intelligence
& 5.809 (.985)
& 6.213 (.572)
& 5.639 (1.059)
& 5.543 (.905)
& 5.66 & $.018^{*}$ & .018 & $.027^{*}$ \\

Agency-oriented anthropomorphism
& 5.769 (2.099)
& 6.820 (1.889)
& 6.219 (1.889)
& 5.721 (2.042)
& 11.56 & $<.001^{***}$ & .037 & $.002^{**}$ \\

Task-performance trust
& 6.698 (1.860)
& 7.429 (1.807)
& 6.734 (1.997)
& 6.573 (1.867)
& 4.20 & $.041^{*}$ & .014 & $.041^{*}$ \\
\addlinespace

Tastes and interests
& 3.213 (.693)
& 3.382 (.574)
& 3.256 (.636)
& 3.317 (.563)
& 0.58 & .447 & .002 & .447 \\

Attitudes and opinions
& 2.650 (.915)
& 2.885 (.790)
& 2.750 (.791)
& 2.807 (.743)
& 0.90 & .344 & .003 & .412 \\

Work or studies
& 2.904 (.785)
& 3.147 (.651)
& 3.032 (.713)
& 3.081 (.702)
& 1.39 & .240 & .005 & .360 \\

Economic and social status
& 2.421 (.873)
& 2.713 (.809)
& 2.545 (.771)
& 2.512 (.719)
& 3.17 & .076 & .010 & .360 \\

Interpersonal relationships
& 2.673 (.806)
& 2.865 (.745)
& 2.778 (.700)
& 2.762 (.699)
& 1.51 & .220 & .005 & .360 \\

Physical appearance
& 2.402 (.868)
& 2.645 (.816)
& 2.511 (.786)
& 2.533 (.646)
& 1.49 & .223 & .005 & .360 \\
\bottomrule
\end{tabular}}

\vspace{2pt}
\begin{minipage}{\textwidth}
\footnotesize
\textit{Note.} The table reports only the focal rendering-by-source-condition
interaction from each Type III 2$\times$2 ANOVA; $df_1=1$ and $df_2=302$.
Partial $\eta_p^2$ is the standardized omnibus effect size. Marginal main effects are omitted from the table for brevity; the
perceived-intelligence source-condition effect reported in the text was
estimated from the same Type III model. Planned simple
rendering contrasts and their standardized effect sizes (Hedges' $g$) are
reported in the text. Perceived quality was the single focal outcome and was
not multiplicity-adjusted. Benjamini--Hochberg correction was applied
separately across the three system-appraisal interactions and the six
disclosure-domain interactions. $^{***}p<.001$, $^{**}p<.01$,
$^{*}p<.05$.
\end{minipage}
\end{table*}

\subsection{How the Appraisals Are Organized: The Structural Model}

\begin{figure*}[htbp]
    \centering
    \includegraphics[width=\textwidth]{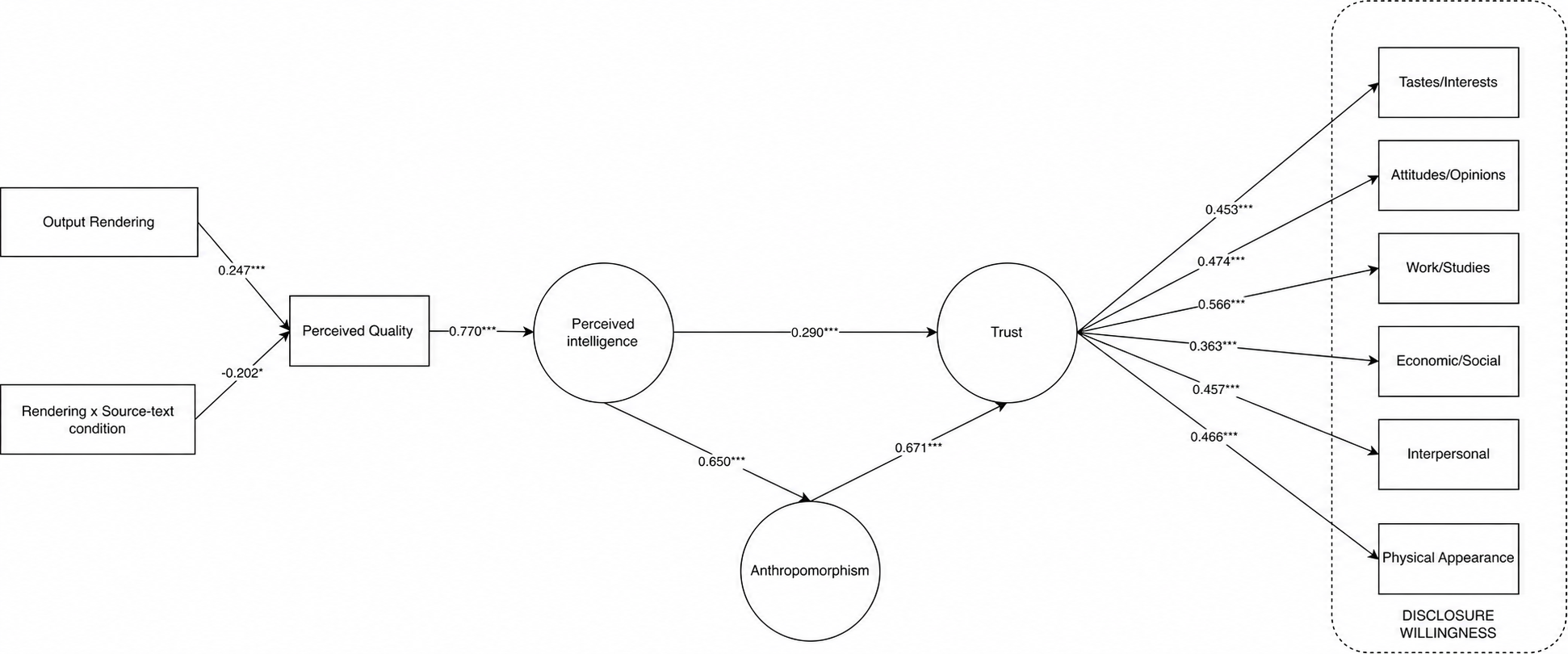}
    \caption{\textbf{Standardized path estimates from the appraisal-structure
model.} The condition-to-quality paths are included to present the complete
structural specification; RQ1a and RQ1b were formally evaluated using the
factorial analyses reported in Section~\ref{sec:quality-effects}. Perceived
intelligence, anthropomorphism, and trust were modeled as latent constructs
(circles). The six disclosure domains were modeled as separate observed
domain-mean outcomes (rectangles). The nonsignificant source-condition path
is omitted for visual clarity but is reported in Table~\ref{tab:modelb}.
 Values shown are standardized coefficients. $***p<.001$, $**p<.01$, $*p<.05$.}
\Description{Structural path diagram. Output rendering and its interaction
with source-text condition point to perceived quality. Perceived quality
points to perceived intelligence. Perceived intelligence points to
anthropomorphism and trust, and anthropomorphism also points to trust. Trust
points to six separate disclosure-willingness domains.}
    \label{fig:sem-results}
\end{figure*}

The factorial analyses above tested whether the observed outcome scores
differed across the study conditions. We next estimated the appraisal-structure
SEM to characterize the theory-ordered concurrent associations among perceived
quality, perceived intelligence, anthropomorphism, task-performance trust, and
stated disclosure willingness. Perceived quality remained an observed
composite, whereas perceived intelligence, anthropomorphism, and trust were
modeled as latent factors. Figure~\ref{fig:sem-results} presents the complete
structural specification.

The condition predictors were retained in the perceived-quality equation to
present the complete structural specification. RQ1a and RQ1b were formally
evaluated using the factorial analyses in
Section~\ref{sec:quality-effects}. The corresponding coefficients in
Table~\ref{tab:modelb} are included for completeness and are not interpreted
as independent confirmatory tests of the research questions. Because the
condition predictors were dummy-coded, the rendering and source-condition
coefficients are conditional on the reference level of the other factor rather
than marginal main effects.

The appraisal-structure model showed acceptable fit, robust $\chi^2(254)=516.04$, $p<.001$, CFI $=.940$, TLI $=.930$, RMSEA $=.058$, 90\% CI $[.051,.065]$, and SRMR $=.044$.

\begin{table*}[htbp]
\centering
\begin{tabular}{lrrrlr}
\toprule
\textbf{Path} & \textbf{$b$} & \textbf{$SE$} & \textbf{$\beta$} & \textbf{95\% CI} & \textbf{$p$} \\
\midrule
Output rendering $\rightarrow$ Perceived quality              & .584 & .169 & .247 & $[.254, .915]$ & $<.001$ \\
Source-text condition  $\rightarrow$ Perceived quality               & $-$.059 & .183 & $-$.025 & $[-.416, .299]$ & .749 \\
Rendering $\times$ source condition  $\rightarrow$ Perceived quality & $-$.566 & .260 & $-$.202 & $[-1.076, -.056]$ & .030 \\
Perceived quality $\rightarrow$ Perceived intelligence        & 1.025 & .095 & .770 & $[.839, 1.210]$ & $<.001$ \\
Perceived intelligence $\rightarrow$ Anthropomorphism         & .546 & .062 & .650 & $[.425, .666]$ & $<.001$ \\
Perceived intelligence $\rightarrow$ Trust                    & .400 & .101 & .290 & $[.202, .599]$ & $<.001$ \\
Anthropomorphism $\rightarrow$ Trust                          & 1.106 & .171 & .671 & $[.771, 1.442]$ & $<.001$ \\
\addlinespace
Trust $\rightarrow$ Tastes and interests                      & .130 & .019 & .453 & $[.092, .168]$ & $<.001$ \\
Trust $\rightarrow$ Attitudes and opinions                    & .178 & .023 & .474 & $[.133, .223]$ & $<.001$ \\
Trust $\rightarrow$ Work or studies                           & .187 & .020 & .566 & $[.148, .227]$ & $<.001$ \\
Trust $\rightarrow$ Economic and social status                & .133 & .024 & .363 & $[.087, .180]$ & $<.001$ \\
Trust $\rightarrow$ Interpersonal relationships               & .155 & .021 & .457 & $[.114, .197]$ & $<.001$ \\
Trust $\rightarrow$ Physical appearance                       & .169 & .022 & .466 & $[.126, .212]$ & $<.001$ \\
\bottomrule
\end{tabular}
\caption{Appraisal-structure model. Reported $p$ values are unadjusted; all six trust--disclosure associations remained significant after Benjamini--Hochberg FDR correction (all FDR-adjusted $p<.001$).}
\label{tab:modelb}
\end{table*}

\subsubsection{Associations among Output and System Appraisals (H1--H4)}

Perceived quality was positively associated with perceived intelligence (H1: $b=1.025$, $\beta=.770$, $p<.001$). Perceived intelligence was positively associated with anthropomorphism (H2: $b=.546$, $\beta=.650$, $p<.001$) and trust (H4: $b=.400$, $\beta=.290$, $p<.001$). Anthropomorphism was also positively associated with trust (H3: $b=1.106$, $\beta=.671$, $p<.001$).

\subsubsection{Disclosure Willingness Across Six Domains (H5a--H5f)}

The disclosure submodel modeled trust as the proximal predictor of disclosure willingness across six personal domains. Trust was positively associated with disclosure willingness in all six: tastes and interests (H5a: $\beta=.453$), attitudes and opinions (H5b: $\beta=.474$), work or studies (H5c: $\beta=.566$), economic and social status (H5d: $\beta=.363$), interpersonal relationships (H5e: $\beta=.457$), and physical appearance (H5f: $\beta=.466$), all $p<.001$ and all significant after Benjamini--Hochberg FDR correction.

The indirect association between perceived anthropomorphism and disclosure willingness through trust was significant in all six domains ($\beta=.244$--$.380$; all 5{,}000-sample percentile-bootstrap confidence intervals excluded zero). Adding direct anthropomorphism-to-disclosure paths did not improve model fit, robust $\Delta\chi^2(6)=3.67$, $p=.722$, and none of the direct paths was individually significant. These results support retaining the more parsimonious statistical specification.

The appraisal-structure model yielded small model-implied indirect associations between rendering and disclosure willingness for the simple texts (\(\beta=.050\)--\(.078\); all bootstrap confidence intervals excluded zero; all FDR-adjusted \(p\) values were .004). The corresponding associations were negligible for the complex texts
($\beta=.009$--$.014$; all confidence intervals included zero; all
FDR-adjusted $p$ values were .927), consistent with the near-zero
model-implied rendering contrast for perceived quality in the complex-source
condition. This contrast is the sum of the rendering and interaction
coefficients in Table~\ref{tab:modelb},
$b=.584+(-.566)=.018$. These conditional indirect associations do not imply a reliable total condition difference in disclosure, which was not detected in the factorial analyses. Because the appraisal variables were measured concurrently and rendering was compared across collection periods, these estimates describe model-implied associations rather than causal mediation.

\section{Discussion}
\subsection{The Evaluability Gap in Source-Visible Translation}

For the simple narratives, ratings followed the source-retention ordering documented by the stimulus audit: fidelity-oriented outputs received higher ratings. For the complex literary-philosophical prose, no rendering difference was detected, although the audit indicated greater source similarity for fidelity-oriented outputs. We refer to this mismatch as an evaluability gap: the source was visible, but overall-quality ratings did not reflect the difference in retained content for the complex stimuli.

Earlier machine-translation research identified a related gap between fluency and adequacy. Users react strongly to disfluent output but may overlook adequacy errors in fluent translations \cite{martindale_fluency_2018}. Sentence-level judgments can also compress differences that become visible when evaluators see document context or inspect errors directly \cite{Laubli2018HumanParity,FischerLaubli2020Difference}. Our finding extends this concern to readability-oriented rewriting in a source-visible interface. Showing the source made evidence available, but did not ensure that the holistic rating reflected the audited retention difference for the complex stimuli. This distinction between access and evaluability is the central HCI contribution.

Several explanations remain plausible. Comparing a difficult source with its output requires effort, which may increase reliance on ease of reading \cite{AlterOppenheimer2009Fluency,ReberSchwarz1999FluencyTruth}. Participants may instead have changed the weight assigned to readability and source retention, or the overall-quality item may have combined dimensions that should have been measured separately. The source-text conditions also differed in genre, provenance, period style, topic, and length. Separating these explanations requires matched stimuli and distinct measures of processing effort, readability, and adequacy.

The source-dependent pattern also appeared in system appraisals (RQ2). For the simple texts, the fidelity-oriented outputs — which the stimulus audit showed retained more source content — received higher ratings of perceived intelligence, agency-oriented attribution, and task-performance trust; for the complex texts, the same audited retention advantage was accompanied by no reliable differences in any of the three appraisals, and all three rendering-by-source-condition interactions were reliable.

\subsection{Output Appraisals and Agency-Oriented Attribution}

Perceived output quality was associated with perceived intelligence ($\beta=.770$, $p<.001$), and perceived intelligence was associated with agency-oriented anthropomorphic attribution ($\beta=.650$, $p<.001$).

Anthropomorphism research has often examined systems that signal humanness through appearance or interaction design \cite{nowak_influence_2005,kim_anthropomorphic_2023,salem_err_2013}. TransLingo omitted those overt cues, although its brand, prototype framing, and task instructions could still shape expectations. In this plain-text utility, ratings of understanding, anticipation, and planning nevertheless covaried with perceived intelligence. A cue-present comparison would help determine how explicit social design changes this relationship.

Elicited agent knowledge offers one interpretation: behavior associated with human agents can activate a human-oriented reading of a nonhuman system \cite{epley_seeing_2007}. The items in this study capture cognitive agency rather than emotional or relational humanness, and effectance and sociality motivations were not measured. The ordering is therefore best read as a theory-guided account of concurrent judgments.

Perceived intelligence and agency-oriented attribution were both associated with task-performance trust ($\beta=.290$ and $\beta=.671$, respectively; both $p<.001$), while attribution and trust remained closely related (HTMT $=.871$). The unity-constraint test supported treating them as distinct constructs, but their overlap argues against reading them as sharply separated stages. The measures describe a connected appraisal structure around this plain-text translation interface.

\subsection{Trust and Disclosure Willingness}

Task-performance trust was the construct most directly associated with stated disclosure willingness in the retained model ($\beta=.363$--$.566$, all FDR-adjusted $p<.001$). The exploratory addition of direct agency-attribution paths did not improve fit, robust $\Delta\chi^2(6)=3.67$, $p=.722$. Prior studies range from users' responses to virtual agents in a disclosure context \cite{mendes_evaluating_2024}, to disclosure intentions in chatbot experiments \cite{fu_are_2026}, to information revealed in a field interaction \cite{xu_identity_disclosure_2024}. Within our concurrent self-reports, the model without direct attribution-to-willingness paths was more parsimonious.

Participants reported willingness rather than disclosing personal information. In these data, task-performance trust was the proximal statistical correlate of stated willingness, consistent with evidence that trust may not translate into actual chatbot disclosure \cite{zhang_zhu_actual_disclosure_2026}.

Output evaluation and information entrustment occur in the same interface, yet our results distinguish them. For the simple texts, rendering contrasts appeared in quality and system appraisals but not in disclosure willingness. Interfaces should therefore support judging what an output preserved separately from deciding whether to submit personal text.

\subsection{Implications for Design and Evaluation}

The evaluability gap may also arise in other source-restating tasks. Scientific summaries can be easy to read while extending claims beyond source evidence \cite{PetersChinYee2025GeneralizationBias}, and explanations can raise confidence without improving accuracy assessment \cite{SteyversEtAl2025LLMKnowledge}. These tasks need not share one psychological mechanism, but they raise a related design question: how can an interface help users inspect what a fluent output preserved or altered? For translation, this motivates testing source--output alignment, omission or simplification indicators, and separate readability and fidelity assessments.

The results also raise a caution for human-in-the-loop evaluation and preference-based optimization of LLMs in translation and other source-restating tasks. If evaluators repeatedly select the output that feels better overall without separately examining source retention, the resulting feedback may favor readable outputs even when they omit or alter source content. In our own stimulus audit, reference-based metrics (BERTScore, COMET) registered the complex-condition retention difference that participants' overall ratings did not, illustrating that automatic source-based measures can remain sensitive where holistic human judgments are not. Evaluation pipelines could therefore test separate readability and fidelity judgments, stratification by source difficulty, and source-based metrics or error-focused review alongside overall preference.

Translation performance and personal-data entrustment require different design support. Task-performance trust — a construct concerning accuracy and reliability rather than privacy or data handling — was the appraisal most closely associated with stated disclosure willingness, and no reliable rendering contrast emerged for disclosure itself. This separation motivates testing explicit information about data retention, model-training use, and third-party sharing rather than expecting users to infer data safety from translation quality or general trust in the system.

\section{Limitations and Future Work}

The sample comprised English-speaking participants who evaluated the system through back-translation. They could compare the English source with the final English output, unlike users who rely on translation because they cannot read one side of the language pair. The fidelity-oriented rendering received higher ratings only for the simple sources; no rendering difference was detected for the complex sources. This design does not establish how the pattern would change when users cannot evaluate the target language. Follow-up work should test that comparison directly.

Source complexity was represented by only two texts at each level and was confounded with genre, provenance, period style, and length. The simple sources were contemporary generated narratives, whereas the complex sources were literary-philosophical prose. The observed interaction therefore applies to these stimuli and does not isolate which text properties produced it. A stronger test would vary complexity across a larger set of texts matched on these other properties. Replication with independently generated system outputs would test whether the pattern persists beyond these prepared renderings.

Data collection spanned August 2023 to February 2025, a period in which baseline familiarity with generative AI was changing. Prior AI familiarity was not measured. Rendering condition was aligned with the data-collection period and therefore was not independently randomized across periods; period-related variation is treated as a design boundary rather than as a temporal finding.

All appraisal constructs were collected in the same post-survey, and anthropomorphism and trust were closely related. Common-method effects may therefore contribute to their associations, and the SEM cannot establish their temporal ordering. Repeated-interaction studies should examine how these appraisals change after translation errors, correction, and recovery.

Self-disclosure was measured as stated willingness rather than actual disclosure behavior. The study also did not measure how participants perceived the intimacy, valence, or sensitivity of individual topics. Future work should measure these perceptions and observe what information users actually choose to translate.

\section{Conclusion}

This study provides evidence of an evaluability gap in the tested AI translation setting. Fidelity-oriented rendering received higher ratings of perceived quality, intelligence, agency-oriented attribution, and task-performance trust for the simple sources, but no reliable rendering difference was detected for the complex sources, despite greater source retention under both conditions. In addition, no reliable rendering difference emerged in stated disclosure willingness under either source-text condition. The appraisal-structure model clarified this distinction: participants who reported greater task-performance trust also reported greater disclosure willingness, but the model-implied indirect association between rendering and disclosure was small for the simple sources and negligible for the complex sources. The HCI contribution is a distinction between source access and source evaluability: showing the source can leave users without enough support to recognize what an output has preserved. Because translation quality and stated disclosure willingness did not respond to the rendering contrast in the same way, source-linked change support and data-handling information should be evaluated as separate interventions in consequential translation tasks.

\appendix
\section{Measurement Items}
\label{app:measurement-items}

This appendix reports the wording presented to participants. 

\subsection{Perceived Intelligence}

\begin{itemize}

    \item[\textbf{PI1}] The machine translation system (TransLingo) can understand the input text accurately.
    \item[\textbf{PI2}] The machine translation system (TransLingo) can provide translations in an understandable manner.
    \item[\textbf{PI3}] The machine translation system (TransLingo) can find and process the necessary information for accurate translations.
    \item[\textbf{PI4}] The machine translation system (TransLingo) is able to provide me with a useful and accurate translation.
\end{itemize}

\textbf{Answer options:} 1 = Strongly disagree, to 7 = Strongly agree.

\subsection{Anthropomorphism}

\begin{itemize}
    \item[\textbf{ANT1}] How smart does this MT system (TransLingo) seem?
    \item[\textbf{ANT2}] How well do you think this machine translation system (TransLingo) can understand the context of the text?
    \item[\textbf{ANT3}] How well do you think this machine translation system (TransLingo) could anticipate what is about to be written, before it is actually written?
    \item[\textbf{ANT4}] How well do you think this machine translation system (TransLingo) can plan and choose the best translation options available?
\end{itemize}

\textbf{Answer options:} 0 = Not at all, to 10 = Very much.

\subsection{Trust}

\begin{itemize}
    \item[\textbf{TR1}] How accurate do you think the average translation is when utilizing this machine translation system (TransLingo)?
    \item[\textbf{TR2}] How much would you trust this machine translation system (TransLingo) to perform accurately in busy, context-rich, real-time language scenarios?
    \item[\textbf{TR3}] How much would you trust this machine translation system (TransLingo) to perform accurately in simpler, less context-heavy language scenarios?
    \item[\textbf{TR4}] How confident would you feel in the translation quality of this system (TransLingo) if you were actively using it?
    \item[\textbf{TR5}] How much do you trust this machine translation system's (TransLingo) ability to provide accurate translations in the next task?
    \item[\textbf{TR6}] How much do you trust this machine translation system's (TransLingo) ability to provide accurate translations throughout this task?
    \item[\textbf{TR7}] How at ease do you feel relying on this machine translation system (TransLingo) for your communication needs?
\end{itemize}

\textbf{Answer options:} 0 = Not at all, to 10 = Very much.

\subsection{Disclosure Willingness}

For each topic listed below, participants were asked:

\begin{quote}
``How comfortable do you feel using TransLingo to translate and disclose this aspect of yourself?''
\end{quote}

\subsubsection{Tastes and Interests}

\begin{itemize}
    \item My favorite foods, the ways I like food prepared.
    \item The kind of music, books, movies, or TV shows that I cannot bear.
    \item My favorite singers, movie stars, writers, places, and brands.
    \item People or organizations that I strongly dislike.
    \item The kind of party or social gathering that I like the best.
    \item The type of social gathering that would bore me or that I would not enjoy.
\end{itemize}

\subsubsection{Attitudes and Opinions}

\begin{itemize}
    \item Political or religious views that I admire.
    \item My indifference or dislikes about certain political or religious views.
    \item Positive comments about the progressing of society.
    \item My criticism about persisting social problems such as poverty and injustice.
    \item Support for certain controversial choices such as dropping out of school.
    \item Skepticism about the choices that people make, e.g., career choices and marriage.
\end{itemize}

\subsubsection{Work or Studies}

\begin{itemize}
    \item What I enjoy and get the most satisfaction from in my present work or study.
    \item Pressures and strains in my work or study.
    \item What I feel are my special strong points for my work or study.
    \item What I find to be the most boring and unenjoyable in work or study.
    \item Things that I accomplish or achieved in work or study.
    \item Frustrations about how my work or study is not being valued at all.
\end{itemize}

\subsubsection{Economic and Social Status}

\begin{itemize}
    \item Items or signals about how wealthy I am, e.g., luxury trips and accessories.
    \item Pressing need for money right now, e.g., outstanding bills and debts.
    \item Optimism about my future financial worth, e.g., getting job offers.
    \item Pessimistic views about my own future employment prospects and salaries.
    \item Connections to people who have high social status.
    \item Feelings of inferiority economically or socially compared with others around me.
\end{itemize}

\subsubsection{Interpersonal Relations and Self-Concept}

\begin{itemize}
    \item Having good times with my significant other.
    \item Disappointments or bad experiences I have had in a romantic relationship.
    \item Missing home or old friends.
    \item Things that I dislike or resent about my friends.
    \item Things that make me especially proud of myself.
    \item What I dislike the most about myself.
\end{itemize}

\subsubsection{Physical Appearance and Sex}

\begin{itemize}
    \item Confidence in my sexual adequacy.
    \item Disappointments in past sexual activity or fear of the first sexual experience.
    \item My standards about the attractiveness of an ideal partner.
    \item Things I do not like about my appearance---nose, eyes, hair, skin, etc.
    \item Happiness about being fit, healthy, and attractive.
    \item Efforts to improve my physical appearance, such as exercise, diet, or surgery.
\end{itemize}

\textbf{Answer options:} 1 = Very uncomfortable; 2 = Somewhat uncomfortable; 3 = Somewhat comfortable; 4 = Very comfortable.

\bibliographystyle{ACM-Reference-Format}
\bibliography{references}

@article{metzger_privacy_2004,
	title = {Privacy, {Trust}, and {Disclosure}: {Exploring} {Barriers} to {Electronic} {Commerce}},
	volume = {9},
	issn = {1083-6101},
	shorttitle = {Privacy, {Trust}, and {Disclosure}},
	url = {https://doi.org/10.1111/j.1083-6101.2004.tb00292.x},
	doi = {10.1111/j.1083-6101.2004.tb00292.x},
	number = {4},
	urldate = {2026-08-07},
	journal = {Journal of Computer-Mediated Communication},
	author = {Metzger, Miriam J.},
	month = jul,
	year = {2004},
	pages = {JCMC942},
}

@article{chen_thinking_2025,
	title = {Thinking twice and weighing outcomes: comparing the dual influence of intelligent healthcare features on word-of-mouth},
	volume = {0},
	issn = {0144-929X},
	shorttitle = {Thinking twice and weighing outcomes},
	url = {https://doi.org/10.1080/0144929X.2025.2502478},
	doi = {10.1080/0144929X.2025.2502478},
	number = {0},
	urldate = {2025-09-17},
	journal = {Behaviour \& Information Technology},
	publisher = {Taylor \& Francis},
	author = {Chen, Aihui and Zhang, Yini and Feng, Nan},
	year = {2025},
	note = {\_eprint: https://doi.org/10.1080/0144929X.2025.2502478},
	pages = {1--16},
}

@article{gray_dimensions_2007,
	title = {Dimensions of {Mind} {Perception}},
	volume = {315},
	issn = {0036-8075, 1095-9203},
	url = {https://www.science.org/doi/10.1126/science.1134475},
	doi = {10.1126/science.1134475},
	language = {en},
	number = {5812},
	urldate = {2026-08-06},
	journal = {Science},
	author = {Gray, Heather M. and Gray, Kurt and Wegner, Daniel M.},
	month = feb,
	year = {2007},
	pages = {619--619},
}

@inproceedings{barayan_analysing_2025,
	address = {Abu Dhabi, UAE},
	title = {Analysing {Zero}-{Shot} {Readability}-{Controlled} {Sentence} {Simplification}},
	url = {https://aclanthology.org/2025.coling-main.452/},
	urldate = {2026-08-05},
	booktitle = {Proceedings of the 31st {International} {Conference} on {Computational} {Linguistics}},
	publisher = {Association for Computational Linguistics},
	author = {Barayan, Abdullah and Camacho-Collados, Jose and Alva-Manchego, Fernando},
	editor = {Rambow, Owen and Wanner, Leo and Apidianaki, Marianna and Al-Khalifa, Hend and Eugenio, Barbara Di and Schockaert, Steven},
	month = jan,
	year = {2025},
	pages = {6762--6781},
}

@article{krasnova_online_2010,
	title = {Online {Social} {Networks}: {Why} {We} {Disclose}},
	volume = {25},
	copyright = {https://journals.sagepub.com/page/policies/text-and-data-mining-license},
	issn = {0268-3962, 1466-4437},
	shorttitle = {Online {Social} {Networks}},
	url = {https://journals.sagepub.com/doi/10.1057/jit.2010.6},
	doi = {10.1057/jit.2010.6},
	language = {en},
	number = {2},
	urldate = {2026-07-28},
	journal = {Journal of Information Technology},
	author = {Krasnova, Hanna and Spiekermann, Sarah and Koroleva, Ksenia and Hildebrand, Thomas},
	month = jun,
	year = {2010},
	pages = {109--125},
}

@article{dinev_extended_2006,
	title = {An {Extended} {Privacy} {Calculus} {Model} for {E}-{Commerce} {Transactions}},
	volume = {17},
	issn = {1047-7047},
	url = {https://www.jstor.org/stable/23015781},
	number = {1},
	urldate = {2026-07-28},
	journal = {Information Systems Research},
	publisher = {INFORMS},
	author = {Dinev, Tamara and Hart, Paul},
	year = {2006},
	pages = {61--80},
}

@inproceedings{rei2020comet,
	address = {Online},
	title = {{COMET}: {A} {Neural} {Framework} for {MT} {Evaluation}},
	url = {https://aclanthology.org/2020.emnlp-main.213/},
	doi = {10.18653/v1/2020.emnlp-main.213},
	booktitle = {Proceedings of the 2020 {Conference} on {Empirical} {Methods} in {Natural} {Language} {Processing} ({EMNLP})},
	publisher = {Association for Computational Linguistics},
	author = {Rei, Ricardo and Stewart, Craig and Farinha, Ana C. and Lavie, Alon},
	month = nov,
	year = {2020},
	pages = {2685--2702},
}

@misc{Zhang2020BERTScore,
	title = {{BERTScore}: {Evaluating} text generation with {BERT}},
	url = {https://openreview.net/forum?id=SkeHuCVFDr},
	howpublished = {International Conference on Learning Representations (ICLR)},
	author = {Zhang, Tianyi and Kishore, Varsha and Wu, Felix and Weinberger, Kilian Q. and Artzi, Yoav},
	year = {2020},
}

@techreport{kincaid1975derivation,
	address = {Millington, TN},
	type = {Research {Branch} {Report}},
	title = {Derivation of {New} {Readability} {Formulas} ({Automated} {Readability} {Index}, {Fog} {Count} and {Flesch} {Reading} {Ease} {Formula}) for {Navy} {Enlisted} {Personnel}},
	url = {https://eric.ed.gov/?id=ED108134},
	number = {8-75},
	institution = {Naval Technical Training Command, Research Branch},
	author = {Kincaid, J. Peter and Fishburne, Jr., Robert P. and Rogers, Richard L. and Chissom, Brad S.},
	month = feb,
	year = {1975},
}

@article{zis-Rosseel2012Lavaan,
	title = {Lavaan: {An} {R} package for structural equation modeling and more. {Version} 0.5-12 ({BETA})},
	volume = {48},
	doi = {http://doi.org/10.18637/jss.v048.i02},
	number = {2},
	journal = {Journal of Statistical Software},
	author = {Rosseel, Y.},
	year = {2012},
	pages = {1--36},
}

@misc{rennekamp_processing_2012,
	address = {Rochester, NY},
	type = {{SSRN} {Scholarly} {Paper}},
	title = {Processing {Fluency} and {Investors}’ {Reactions} to {Disclosure} {Readability}},
	url = {https://papers.ssrn.com/abstract=1881847},
	doi = {10.2139/ssrn.1881847},
	language = {en},
	urldate = {2026-05-20},
	publisher = {Social Science Research Network},
	author = {Rennekamp, Kristina M.},
	month = feb,
	year = {2012},
}

@misc{han_machine_2018,
	title = {Machine {Translation} {Evaluation} {Resources} and {Methods}: {A} {Survey}},
	language = {en},
	url = {https://arxiv.org/abs/1605.04515},
	author = {Han, Lifeng},
	year = {2018},
	note = {Conference poster presented at IPRC 2018; arXiv:1605.04515},
}

@inproceedings{white_evaluation_1993,
	address = {Stroudsburg, PA, USA},
	title = {Evaluation of {Machine} {Translation}},
	url = {https://aclanthology.org/H93-1040/},
	doi = {10.3115/1075671.1075717},
	urldate = {2026-05-19},
	booktitle = {Human {Language} {Technology}: {Proceedings} of a {Workshop} {Held} at {Plainsboro}, {New} {Jersey}, {March} 21-24, 1993},
	publisher = {Association for Computational Linguistics},
	author = {White, John S. and O'Connell, Theresa A.},
	year = {1993},
	pages = {206--210},
}

@article{fu_are_2026,
	title = {Are {Consumers} {Willing} to {Disclose} {Information} to {AI}-{Based} {Chatbot}? {The} {Roles} of {Anthropomorphism} and {Regulatory} {Focus}},
	volume = {25},
	copyright = {© 2025 John Wiley \& Sons Ltd.},
	issn = {1479-1838},
	shorttitle = {Are {Consumers} {Willing} to {Disclose} {Information} to {AI}-{Based} {Chatbot}?},
	url = {https://onlinelibrary.wiley.com/doi/abs/10.1002/cb.70076},
	doi = {10.1002/cb.70076},
	language = {en},
	number = {1},
	urldate = {2026-05-19},
	journal = {Journal of Consumer Behaviour},
	author = {Fu, Huijian and Shang, Lixia and Lin, Weican and Du, Helen S.},
	year = {2026},
	note = {\_eprint: https://onlinelibrary.wiley.com/doi/pdf/10.1002/cb.70076},
	pages = {394--408},
}

@article{zhang_zhu_actual_disclosure_2026,
	title = {Actual Self-disclosure to Anthropomorphic {AI} Chatbots: A Contextual Privacy Calculus Approach},
	volume = {18},
	number = {1},
	pages = {31--60},
	url = {https://aisel.aisnet.org/thci/vol18/iss1/2/},
	doi = {10.17705/1thci.00237},
	journal = {AIS Transactions on Human-Computer Interaction},
	author = {Zhang, Mingxin and Zhu, Hou},
	year = {2026},
}

@article{xu_identity_disclosure_2024,
	title = {Identity Disclosure and Anthropomorphism in Voice Chatbot Design: A Field Experiment},
	volume = {72},
	number = {1},
	pages = {223--241},
	url = {https://pubsonline.informs.org/doi/10.1287/mnsc.2022.03833},
	doi = {10.1287/mnsc.2022.03833},
	journal = {Management Science},
	author = {Xu, Yuqian and Dai, Hongyan and Yan, Wanfeng},
	year = {2026},
	note = {Published online August 26, 2024},
}

@book{jourard1971self,
	address = {New York, NY, USA},
	title = {Self-disclosure: {An} experimental analysis of the transparent self.},
	publisher = {John Wiley},
	author = {Jourard, Sidney M},
	year = {1971},
}

@inproceedings{mendes_evaluating_2024,
	address = {Piscataway, NJ, USA},
	title = {Evaluating the {Impact} of {Anthropomorphism} and {Verbal} {Interaction} on {Users}’ {Perception} of a {Virtual} {Agent} in {Facilitating} {Self}-{Disclosure}},
	issn = {2573-3060},
	url = {https://ieeexplore.ieee.org/abstract/document/10639572},
	doi = {10.1109/SeGAH61285.2024.10639572},
	urldate = {2026-05-19},
	booktitle = {2024 {IEEE} 12th {International} {Conference} on {Serious} {Games} and {Applications} for {Health} ({SeGAH})},
	publisher = {IEEE},
	author = {Mendes, Diana C. G. and Costa, Rita and Agrela, Gabriel and Badia, Sergi Bermúdez i and Cameirão, Mónica S.},
	month = aug,
	year = {2024},
	note = {ISSN: 2573-3060},
	pages = {1--8},
}

@article{salem_err_2013,
	title = {To {Err} is {Human}(-like): {Effects} of {Robot} {Gesture} on {Perceived} {Anthropomorphism} and {Likability}},
	volume = {5},
	issn = {1875-4805},
	shorttitle = {To {Err} is {Human}(-like)},
	url = {https://doi.org/10.1007/s12369-013-0196-9},
	doi = {10.1007/s12369-013-0196-9},
	language = {en},
	number = {3},
	urldate = {2025-09-23},
	journal = {International Journal of Social Robotics},
	author = {Salem, Maha and Eyssel, Friederike and Rohlfing, Katharina and Kopp, Stefan and Joublin, Frank},
	month = aug,
	year = {2013},
	pages = {313--323},
}

@article{waytz_mind_2014,
	title = {The mind in the machine: {Anthropomorphism} increases trust in an autonomous vehicle},
	volume = {52},
	issn = {0022-1031},
	shorttitle = {The mind in the machine},
	url = {https://www.sciencedirect.com/science/article/pii/S0022103114000067},
	doi = {10.1016/j.jesp.2014.01.005},
	urldate = {2025-09-17},
	journal = {Journal of Experimental Social Psychology},
	author = {Waytz, Adam and Heafner, Joy and Epley, Nicholas},
	month = may,
	year = {2014},
	pages = {113--117},
}

@article{nowak_influence_2005,
	title = {The {Influence} of the {Avatar} on {Online} {Perceptions} of {Anthropomorphism}, {Androgyny}, {Credibility}, {Homophily}, and {Attraction}},
	volume = {11},
	issn = {1083-6101},
	url = {https://onlinelibrary.wiley.com/doi/abs/10.1111/j.1083-6101.2006.tb00308.x},
	doi = {10.1111/j.1083-6101.2006.tb00308.x},
	language = {en},
	number = {1},
	urldate = {2025-09-23},
	journal = {Journal of Computer-Mediated Communication},
	author = {Nowak, Kristine L. and Rauh, Christian},
	year = {2005},
	note = {\_eprint: https://onlinelibrary.wiley.com/doi/pdf/10.1111/j.1083-6101.2006.tb00308.x},
	pages = {153--178},
}

@inproceedings{chiesurin_dangers_2023,
	address = {Toronto, Canada},
	title = {The {Dangers} of trusting {Stochastic} {Parrots}: {Faithfulness} and {Trust} in {Open}-domain {Conversational} {Question} {Answering}},
	shorttitle = {The {Dangers} of trusting {Stochastic} {Parrots}},
	url = {https://aclanthology.org/2023.findings-acl.60},
	doi = {10.18653/v1/2023.findings-acl.60},
	language = {en},
	urldate = {2026-05-16},
	booktitle = {Findings of the {Association} for {Computational} {Linguistics}: {ACL} 2023},
	publisher = {Association for Computational Linguistics},
	author = {Chiesurin, Sabrina and Dimakopoulos, Dimitris and Sobrevilla Cabezudo, Marco Antonio and Eshghi, Arash and Papaioannou, Ioannis and Rieser, Verena and Konstas, Ioannis},
	year = {2023},
	pages = {947--959},
}

@inproceedings{Moussawi_Koufaris_2019,
	address = {Honolulu, HI, USA},
	title = {Perceived intelligence and perceived anthropomorphism of personal intelligent agents: {Scale} development and validation},
	doi = {10.24251/hicss.2019.015},
	booktitle = {Proceedings of the 52nd Hawaii International Conference on System Sciences},
	publisher = {Hawaii International Conference on System Sciences},
	pages = {115--124},
	author = {Moussawi, Sara and Koufaris, Marios},
	month = jan,
	year = {2019},
}

@article{epley_seeing_2007,
	address = {US},
	title = {On seeing human: {A} three-factor theory of anthropomorphism},
	volume = {114},
	issn = {1939-1471},
	shorttitle = {On seeing human},
	doi = {10.1037/0033-295X.114.4.864},
	number = {4},
	journal = {Psychological Review},
	publisher = {American Psychological Association},
	author = {Epley, Nicholas and Waytz, Adam and Cacioppo, John T.},
	year = {2007},
	pages = {864--886},
}

@article{nass_machines_2000,
	title = {Machines and mindlessness: {Social} responses to computers},
	volume = {56},
	url = {https://spssi.onlinelibrary.wiley.com/doi/abs/10.1111/0022-4537.00153},
	doi = {https://doi.org/10.1111/0022-4537.00153},
	number = {1},
	journal = {Journal of Social Issues},
	author = {Nass, Clifford and Moon, Youngme},
	year = {2000},
	note = {tex.eprint: https://spssi.onlinelibrary.wiley.com/doi/pdf/10.1111/0022-4537.00153},
	pages = {81--103},
}

@article{moussawi_how_2021,
	title = {How perceptions of intelligence and anthropomorphism affect adoption of personal intelligent agents},
	volume = {31},
	issn = {1422-8890},
	url = {https://doi.org/10.1007/s12525-020-00411-w},
	doi = {10.1007/s12525-020-00411-w},
	language = {en},
	number = {2},
	urldate = {2025-09-24},
	journal = {Electronic Markets},
	author = {Moussawi, Sara and Koufaris, Marios and Benbunan-Fich, Raquel},
	month = jun,
	year = {2021},
	pages = {343--364},
}

@inproceedings{martindale_fluency_2018,
	address = {Boston, MA},
	title = {Fluency {Over} {Adequacy}: {A} {Pilot} {Study} in {Measuring} {User} {Trust} in {Imperfect} {MT}},
	shorttitle = {Fluency {Over} {Adequacy}},
	url = {https://aclanthology.org/W18-1803/},
	urldate = {2026-05-16},
	booktitle = {Proceedings of the 13th {Conference} of the {Association} for {Machine} {Translation} in the {Americas} ({Volume} 1: {Research} {Track})},
	publisher = {Association for Machine Translation in the Americas},
	author = {Martindale, Marianna and Carpuat, Marine},
	editor = {Cherry, Colin and Neubig, Graham},
	month = mar,
	year = {2018},
	pages = {13--25},
}

@inproceedings{natarajan_effects_2020,
	address = {Cambridge, United Kingdom},
	series = {Hri '20},
	title = {Effects of anthropomorphism and accountability on trust in human robot interaction},
	isbn = {978-1-4503-6746-2},
	url = {https://doi.org/10.1145/3319502.3374839},
	doi = {10.1145/3319502.3374839},
	booktitle = {Proceedings of the 2020 {ACM}/{IEEE} international conference on human-robot interaction},
	publisher = {Association for Computing Machinery},
	author = {Natarajan, Manisha and Gombolay, Matthew},
	year = {2020},
	note = {Number of pages: 10
tex.address: New York, NY, USA},
	pages = {33--42},
}

@article{troshani_we_2021,
	title = {Do we trust in {AI}? {Role} of anthropomorphism and intelligence},
	volume = {61},
	url = {https://doi.org/10.1080/08874417.2020.1788473},
	doi = {10.1080/08874417.2020.1788473},
	number = {5},
	journal = {Journal of Computer Information Systems},
	publisher = {Taylor \& Francis},
	author = {Troshani, Indrit and Hill, Sally Rao and Sherman, Claire and Arthur, Damien},
	year = {2021},
	note = {tex.eprint: https://doi.org/10.1080/08874417.2020.1788473},
	pages = {481--491},
}

@article{kim_anthropomorphic_2023,
	title = {Anthropomorphic response: {Understanding} interactions between humans and artificial intelligence agents},
	volume = {139},
	issn = {0747-5632},
	shorttitle = {Anthropomorphic response},
	url = {https://www.sciencedirect.com/science/article/pii/S0747563222003326},
	doi = {10.1016/j.chb.2022.107512},
	urldate = {2025-09-18},
	journal = {Computers in Human Behavior},
	author = {Kim, Joohee and Im, Il},
	month = feb,
	year = {2023},
	pages = {107512},
}

@article{AlterOppenheimer2009Fluency,
	title = {Uniting the tribes of fluency to form a metacognitive nation},
	volume = {13},
	doi = {10.1177/1088868309341564},
	number = {3},
	journal = {Personality and Social Psychology Review},
	author = {Alter, Adam L. and Oppenheimer, Daniel M.},
	year = {2009},
	pages = {219--235},
}

@article{ReberSchwarz1999FluencyTruth,
	title = {Effects of perceptual fluency on judgments of truth},
	volume = {8},
	doi = {10.1006/ccog.1999.0386},
	number = {3},
	journal = {Consciousness and Cognition},
	author = {Reber, Rolf and Schwarz, Norbert},
	year = {1999},
	pages = {338--342},
}

@article{PetersChinYee2025GeneralizationBias,
	title = {Generalization bias in large language model summarization of scientific research},
	volume = {12},
	url = {https://doi.org/10.1098/rsos.241776},
	doi = {10.1098/rsos.241776},
	journal = {Royal Society Open Science},
	author = {Peters, Uwe and Chin-Yee, Benjamin},
	year = {2025},
	pages = {241776},
}

@article{SteyversEtAl2025LLMKnowledge,
	title = {What large language models know and what people think they know},
	volume = {7},
	url = {https://doi.org/10.1038/s42256-024-00976-7},
	doi = {10.1038/s42256-024-00976-7},
	journal = {Nature Machine Intelligence},
	author = {Steyvers, Mark and Tejeda, Heliodoro and Kumar, Aakriti and Belem, Catarina and Karny, Sheer and Hu, Xinyue and Mayer, Lukas W. and Smyth, Padhraic},
	year = {2025},
	pages = {221--231},
}

@article{HancockNaamanLevy2020AIMC,
	title = {{AI}-mediated communication: {Definition}, research agenda, and ethical considerations},
	volume = {25},
	url = {https://doi.org/10.1093/jcmc/zmz022},
	doi = {10.1093/jcmc/zmz022},
	number = {1},
	journal = {Journal of Computer-Mediated Communication},
	author = {Hancock, Jeffrey T. and Naaman, Mor and Levy, Karen},
	year = {2020},
	pages = {89--100},
}

@inproceedings{FischerLaubli2020Difference,
	address = {Lisboa, Portugal},
	title = {What's the difference between professional human and machine translation? {A} blind multi-language study on domain-specific {MT}},
	url = {https://aclanthology.org/2020.eamt-1.23/},
	booktitle = {Proceedings of the 22nd annual conference of the european association for machine translation},
	publisher = {European Association for Machine Translation},
	author = {Fischer, Lukas and Läubli, Samuel},
	year = {2020},
	pages = {215--224},
}

@inproceedings{Laubli2018HumanParity,
	address = {Brussels, Belgium},
	title = {Has machine translation achieved human parity? {A} case for document-level evaluation},
	url = {https://aclanthology.org/D18-1512/},
	doi = {10.18653/v1/D18-1512},
	booktitle = {Proceedings of the 2018 conference on empirical methods in natural language processing},
	publisher = {Association for Computational Linguistics},
	author = {Läubli, Samuel and Sennrich, Rico and Volk, Martin},
	year = {2018},
	pages = {4791--4796},
}

@inproceedings{10.1145/2858036.2858414,
	address = {San Jose, California, USA},
	series = {Chi '16},
	title = {Anonymity, intimacy and self-disclosure in social media},
	isbn = {978-1-4503-3362-7},
	url = {https://doi.org/10.1145/2858036.2858414},
	doi = {10.1145/2858036.2858414},
	booktitle = {Proceedings of the 2016 {CHI} conference on human factors in computing systems},
	publisher = {Association for Computing Machinery},
	author = {Ma, Xiao and Hancock, Jeff and Naaman, Mor},
	year = {2016},
	note = {Number of pages: 13
tex.address: New York, NY, USA},
	pages = {3857--3869},
}

@article{Jourard_Lasakow_1958,
	title = {Some factors in self-disclosure},
	volume = {56},
	doi = {10.1037/h0043357},
	number = {1},
	journal = {The Journal of Abnormal and Social Psychology},
	publisher = {American Psychological Association},
	author = {Jourard, Sidney M. and Lasakow, Paul},
	year = {1958},
	pages = {91--98},
}

@inproceedings{pascoal_improving_2026,
	address = {Cham},
	title = {Improving {Text} {Readability} to {Support} {Student} {Comprehension} and {Learning}: {An} {LLM}-{Powered} {Approach}},
	isbn = {978-3-032-03870-8},
	shorttitle = {Improving {Text} {Readability} to {Support} {Student} {Comprehension} and {Learning}},
	doi = {10.1007/978-3-032-03870-8_20},
	language = {en},
	booktitle = {Two {Decades} of {TEL}. {From} {Lessons} {Learnt} to {Challenges} {Ahead}},
	publisher = {Springer Nature Switzerland},
	author = {Pascoal, Guilherme and van den Bosch, Marlinde and Viberg, Olga and Wong, Jacqueline and Epp, Carrie Demmans},
	editor = {Tammets, Kairit and Sosnovsky, Sergey and Ferreira Mello, Rafael and Pishtari, Gerti and Nazaretsky, Tanya},
	year = {2026},
	pages = {291--305},
}

@article{ataman_machine_2025,
	title = {Machine {Translation} in the {Era} of {Large} {Language} {Models}:{A} {Survey} of {Historical} and {Emerging} {Problems}},
	volume = {16},
	copyright = {http://creativecommons.org/licenses/by/3.0/},
	issn = {2078-2489},
	shorttitle = {Machine {Translation} in the {Era} of {Large} {Language} {Models}},
	url = {https://www.mdpi.com/2078-2489/16/9/723},
	doi = {10.3390/info16090723},
	language = {en},
	number = {9},
	urldate = {2026-05-19},
	journal = {Information},
	publisher = {Multidisciplinary Digital Publishing Institute},
	author = {Ataman, Duygu and Birch, Alexandra and Habash, Nizar and Federico, Marcello and Koehn, Philipp and Cho, Kyunghyun},
	month = sep,
	year = {2025},
	pages = {723},
}

@article{bindels_machine_2024,
	title = {Machine {Translation} in the {Workplace}: {Deciding} on the {Whether}, {Why} and {How}},
	volume = {11},
	number = {1},
	pages = {Article 4},
	url = {https://commons.emich.edu/gabc/vol11/iss1/4/},
	language = {en},
	journal = {Global Advances in Business Communication},
	author = {Bindels, Joop and Dorst, Aletta G and Pluymaekers, Mark},
	year = {2024},
}

@article{sahin_multilingual_2014,
	title = {Multilingual {Chat} through {Machine} {Translation}: {A} {Case} of {English}-{Russian}},
	volume = {58},
	issn = {1492-1421, 0026-0452},
	shorttitle = {Multilingual {Chat} through {Machine} {Translation}},
	url = {http://id.erudit.org/iderudit/1024180ar},
	doi = {10.7202/1024180ar},
	number = {2},
	urldate = {2026-05-19},
	journal = {Meta},
	author = {Şahin, Mehmet and Duman, Derya},
	month = mar,
	year = {2014},
	pages = {397--410},
}

@article{vieira_understanding_2021,
	title = {Understanding the societal impacts of machine translation: a critical review of the literature on medical and legal use cases},
	volume = {24},
	issn = {1369-118X, 1468-4462},
	shorttitle = {Understanding the societal impacts of machine translation},
	url = {https://www.tandfonline.com/doi/full/10.1080/1369118X.2020.1776370},
	doi = {10.1080/1369118X.2020.1776370},
	language = {en},
	number = {11},
	urldate = {2026-05-19},
	journal = {Information, Communication \& Society},
	author = {Vieira, Lucas Nunes and O’Hagan, Minako and O’Sullivan, Carol},
	month = aug,
	year = {2021},
	pages = {1515--1532},
}

@inproceedings{salimzadeh_missing_2023,
	address = {New York, NY, USA},
	series = {{UMAP} '23},
	title = {A {Missing} {Piece} in the {Puzzle}: {Considering} the {Role} of {Task} {Complexity} in {Human}-{AI} {Decision} {Making}},
	isbn = {978-1-4503-9932-6},
	shorttitle = {A {Missing} {Piece} in the {Puzzle}},
	url = {https://dl.acm.org/doi/10.1145/3565472.3592959},
	doi = {10.1145/3565472.3592959},
	urldate = {2026-03-03},
	booktitle = {Proceedings of the 31st {ACM} {Conference} on {User} {Modeling}, {Adaptation} and {Personalization}},
	publisher = {Association for Computing Machinery},
	author = {Salimzadeh, Sara and He, Gaole and Gadiraju, Ujwal},
	month = jun,
	year = {2023},
	pages = {215--227},
}

@article{moussawi_effect_2021,
	title = {The effect of voice and humour on users’ perceptions of personal intelligent agents},
	volume = {40},
	issn = {0144-929X, 1362-3001},
	url = {https://www.tandfonline.com/doi/full/10.1080/0144929X.2020.1772368},
	doi = {10.1080/0144929X.2020.1772368},
	language = {en},
	number = {15},
	urldate = {2025-11-19},
	journal = {Behaviour \& Information Technology},
	author = {Moussawi, Sara and Benbunan-Fich, Raquel},
	month = nov,
	year = {2021},
	pages = {1603--1626},
}

@article{10.3389/fpsyg.2023.1129294,
	title = {Happiness and high reliability develop affective trust in in-vehicle agents},
	volume = {14},
	issn = {1664-1078},
	url = {https://www.frontiersin.org/journals/psychology/articles/10.3389/fpsyg.2023.1129294},
	doi = {10.3389/fpsyg.2023.1129294},
	journal = {Frontiers in Psychology},
	author = {Zieger, Scott and Dong, Jiayuan and Taylor, Skye and Sanford, Caitlyn and Jeon, Myounghoon},
	year = {2023},
	pages = {1129294},
}

@article{rubin_friendship_1978,
	title = {Friendship, proximity, and self-disclosure},
	volume = {46},
	url = {https://onlinelibrary.wiley.com/doi/abs/10.1111/j.1467-6494.1978.tb00599.x},
	doi = {https://doi.org/10.1111/j.1467-6494.1978.tb00599.x},
	number = {1},
	journal = {Journal of Personality},
	author = {Rubin, Zick and Shenker, Stephen},
	year = {1978},
	note = {tex.eprint: https://onlinelibrary.wiley.com/doi/pdf/10.1111/j.1467-6494.1978.tb00599.x},
	pages = {1--22},
}

@book{chall_readability_1995,
	address = {Brookline, MA, USA},
	title = {Readability {Revisited}: {The} {New} {Dale}-{Chall} {Readability} {Formula}},
	isbn = {978-1-57129-008-3},
	shorttitle = {Readability {Revisited}},
	language = {en},
	publisher = {Brookline Books},
	author = {Chall, Jeanne Sternlicht and Dale, Edgar},
	year = {1995},
	note = {Google-Books-ID: 2nbuAAAAMAAJ},
}

@article{tusseyeva_perceived_2024,
	title = {Perceived {Intelligence} in {Human}-{Robot} {Interaction}: {A} {Review}},
	volume = {12},
	issn = {2169-3536},
	shorttitle = {Perceived {Intelligence} in {Human}-{Robot} {Interaction}},
	url = {https://ieeexplore.ieee.org/document/10714350},
	doi = {10.1109/ACCESS.2024.3478751},
	urldate = {2025-08-11},
	journal = {IEEE Access},
	author = {Tusseyeva, Inara and Sandygulova, Anara and Rubagotti, Matteo},
	year = {2024},
	pages = {151348--151359},
}

@article{lu_role_2025,
	title = {The role of intelligence, trust and interpersonal job characteristics in employees’ {AI} usage acceptance},
	volume = {126},
	issn = {0278-4319},
	url = {https://www.sciencedirect.com/science/article/pii/S027843192400344X},
	doi = {10.1016/j.ijhm.2024.104032},
	urldate = {2025-09-17},
	journal = {International Journal of Hospitality Management},
	author = {Lu, Cheng-Chieh Allan and Yeh, Chu-Chen Rosa and Lai, Chih-Chien Steven},
	month = apr,
	year = {2025},
	pages = {104032},
}

@article{gomes_anthropomorphism_2025,
	title = {Anthropomorphism in artificial intelligence: a game-changer for brand marketing},
	volume = {11},
	issn = {2314-7210},
	shorttitle = {Anthropomorphism in artificial intelligence},
	url = {https://doi.org/10.1186/s43093-025-00423-y},
	doi = {10.1186/s43093-025-00423-y},
	number = {1},
	urldate = {2025-09-18},
	journal = {Future Business Journal},
	author = {Gomes, Sofia and Lopes, João M. and Nogueira, Elisabete},
	month = jan,
	year = {2025},
	pages = {2},
}

@article{chan_moderating_2015,
	title = {The moderating roles of subjective (perceived) and objective task complexity in system use and performance},
	volume = {51},
	issn = {0747-5632},
	url = {https://www.sciencedirect.com/science/article/pii/S0747563215003544},
	doi = {10.1016/j.chb.2015.04.059},
	urldate = {2025-08-07},
	journal = {Computers in Human Behavior},
	author = {Chan, Siew H. and Song, Qian and Yao, Lee J.},
	month = oct,
	year = {2015},
	pages = {393--402},
}

@book{aurelius1923thoughts,
	address = {London, UK},
	series = {New aldine library},
	title = {The thoughts of the emperor marcus aurelius antoninus},
	url = {https://books.google.com/books?id=HLnacDFs30IC},
	publisher = {M. Hopkinson},
	author = {Aurelius, M. and Long, G.},
	year = {1923},
}

@article{ma_effect_2025,
	title = {Effect of anthropomorphism and perceived intelligence in chatbot avatars of visual design on user experience: accounting for perceived empathy and trust},
	volume = {7},
	issn = {2624-9898},
	shorttitle = {Effect of anthropomorphism and perceived intelligence in chatbot avatars of visual design on user experience},
	url = {https://www.frontiersin.org/journals/computer-science/articles/10.3389/fcomp.2025.1531976/full},
	doi = {10.3389/fcomp.2025.1531976},
	language = {English},
	urldate = {2025-09-23},
	journal = {Frontiers in Computer Science},
	publisher = {Frontiers},
	author = {Ma, Ning and Khynevych, Ruslana and Hao, Yunqiang and Wang, Yahui},
	month = may,
	year = {2025},
	pages = {1531976},
}

@inproceedings{koda_agents_1996,
	address = {Piscataway, NJ, USA},
	title = {Agents with faces: the effect of personification},
	shorttitle = {Agents with faces},
	url = {https://ieeexplore.ieee.org/document/568812},
	doi = {10.1109/ROMAN.1996.568812},
	urldate = {2025-09-16},
	booktitle = {Proceedings 5th {IEEE} {International} {Workshop} on {Robot} and {Human} {Communication}. {RO}-{MAN}'96 {TSUKUBA}},
	publisher = {IEEE},
	author = {Koda, T. and Maes, P.},
	month = nov,
	year = {1996},
	pages = {189--194},
}

@article{wood_task_1986,
	title = {Task complexity: {Definition} of the construct},
	volume = {37},
	issn = {0749-5978},
	shorttitle = {Task complexity},
	url = {https://www.sciencedirect.com/science/article/pii/0749597886900440},
	doi = {10.1016/0749-5978(86)90044-0},
	number = {1},
	urldate = {2025-09-16},
	journal = {Organizational Behavior and Human Decision Processes},
	author = {Wood, Robert E},
	month = feb,
	year = {1986},
	pages = {60--82},
}
\end{document}